\documentclass[reprint,prd,letterpaper,superscriptaddress,amsmath,amssymb,aps] {revtex4-1}

\usepackage{graphicx}
\usepackage{dcolumn}
\usepackage{todonotes} 
\usepackage{amsmath} 
\usepackage{bm}
\usepackage{siunitx}
\usepackage{subfigure}
\usepackage{pifont}
\usepackage{xcolor}
\usepackage[implicit=true, debug=false, 
plainpages=false, colorlinks=true, 
linkcolor=black, urlcolor=black, citecolor=black, filecolor=black, 
hyperindex=true]{hyperref}
\AtBeginDocument{\hypersetup{citecolor=black,linkcolor=black,urlcolor=black,filecolor=black}}

\newcommand{\atUC}{\affiliation{Kavli Institute for Cosmological Physics, Dept. of Physics, Enrico Fermi Institute, University of Chicago, Chicago, IL 60637, USA}}
\newcommand{\atUCL}{\affiliation{Dept. of Physics and Astronomy, University College London, London, United Kingdom}}
\newcommand{\atUH}{\affiliation{Dept. of Physics and Astronomy, University of Hawaii, Manoa, HI 96822, USA}}
\newcommand{\atUCLA}{\affiliation{Dept. of Physics and Astronomy, University of California Los Angeles, Los Angeles, CA 90095, USA}}
\newcommand{\atCalPoly}{\affiliation{Physics Dept., California Polytechnic State University, San Luis Obispo, CA 93407, USA}}

\begin{document}

\title{Measurements and Modeling of Near-Surface Radio Propagation in \\ Glacial Ice and Implications 
for Neutrino Experiments}
\author{C.~Deaconu}\atUC
\author{A.~G.~Vieregg}\atUC
\author{S.~A.~Wissel}\atCalPoly
\author{J.~Bowen}\atUC
\author{S.~Chipman}\atUC
\author{A.~Gupta}\atUCL
\author{C.~Miki}\atUH
\author{R.~J.~Nichol}\atUCL
\author{D.~Saltzberg}\atUCLA

\date{\today}

\begin{abstract}
We present measurements of radio transmission in the $\sim100$ MHz range 
through a $\sim100$~m deep region below the surface of the ice at Summit Station,
Greenland, called the firn.  In the firn, the index of refraction 
changes due to the transition from snow at the surface to glacial ice below,
affecting the propagation of radio signals in that region. We compare our observations
to a finite-difference time-domain (FDTD) electromagnetic wave simulation, which
supports the existence of three classes of propagation:  
a bulk propagation ray-bending mode that leads to so-called ``shadowed'' 
regions for certain geometries of transmission, 
a surface-wave mode induced by the ice/air interface, 
and an arbitrary-depth horizontal propagation mode that requires perturbations from a smooth
density gradient.  In the non-shadowed region, 
our measurements are consistent with the bulk propagation ray-bending mode both in timing and in amplitude.  
We also observe signals in the shadowed region, in conflict with a bulk-propagation-only 
ray-bending model, but consistent with FDTD simulations using a variety of firn models
for Summit Station.  The amplitude and timing of our 
measurements in all geometries are consistent with the predictions from FDTD simulations.
In the shadowed region, 
the amplitude of the observed signals is consistent with 
a best-fit coupling fraction value of $2.4$\% (0.06\% in power) or less
to a surface or horizontal propagation mode from the 
bulk propagation mode. 
The relative amplitude of observable signals in the two regions
is important for experiments that aim to detect radio emission from
astrophysical high-energy neutrinos interacting in glacial ice, which rely on a radio propagation model
to inform simulations and perform event reconstruction.
\end{abstract}

\maketitle

\section{Introduction} 
Experiments that aim to detect impulsive radio emission in the $\sim100$~MHz -- 1~GHz range created 
by high-energy astrophysical or cosmogenic neutrinos interacting in glacial ice require an understanding
of radio propagation through glacial ice.
In the bulk ice, this is presumed to be straightforward -- rays propagate along straight lines
in the glacial ice where the ice density and index of refraction are nearly constant.  However, in the 
$\sim100$~m or more below the upper surface of the ice, the snow at the surface slowly transitions
from loose snow to glacial ice below in a region called the firn.  This firn region represents
a density gradient from that of loose snow to that of ice~\cite{arth,koci}, 
corresponding to an index of refraction of $n\approx1.35$ for 
the snow near the surface to $n=1.78$ for bulk ice in the radio frequency range
of interest for neutrino detection~\cite{avva,barwick}.  Furthermore, annual variations
in the firn mean that the gradient is not perfectly smooth -- there are known layers in the firn, 
especially evident near the surface~\cite{hawley}. 
The changing index of refraction causes the paths of electromagnetic waves to bend and scatter, 
affecting the propagation time and distance. To accurately calculate experimental 
sensitivity and reconstruct neutrino energy and arrival direction, the effects of propagation must be included in 
simulations and event reconstruction algorithms of in-ice neutrino experiments.

Ray-tracing models of radio propagation through the firn indicate that rays bend such that there
is a so-called ``shadow'' region, where a receiver placed above a transmitter and horizontally displaced from that 
transmitter will not be able to see emission from the transmitter in certain
geometries~\cite{barwick, araWhitepaper, phasedArray}. 
For neutrino detection, this means that neutrinos that interact in the shadow region of a given receiver cannot 
be seen by that receiver, and implies that a deeper receiver 
will be able to observe a larger instantaneous solid angle compared to a surface or near-surface receiver.  
\begin{figure}[!ht]
      \begin{center}
        \includegraphics[width=8.5cm]{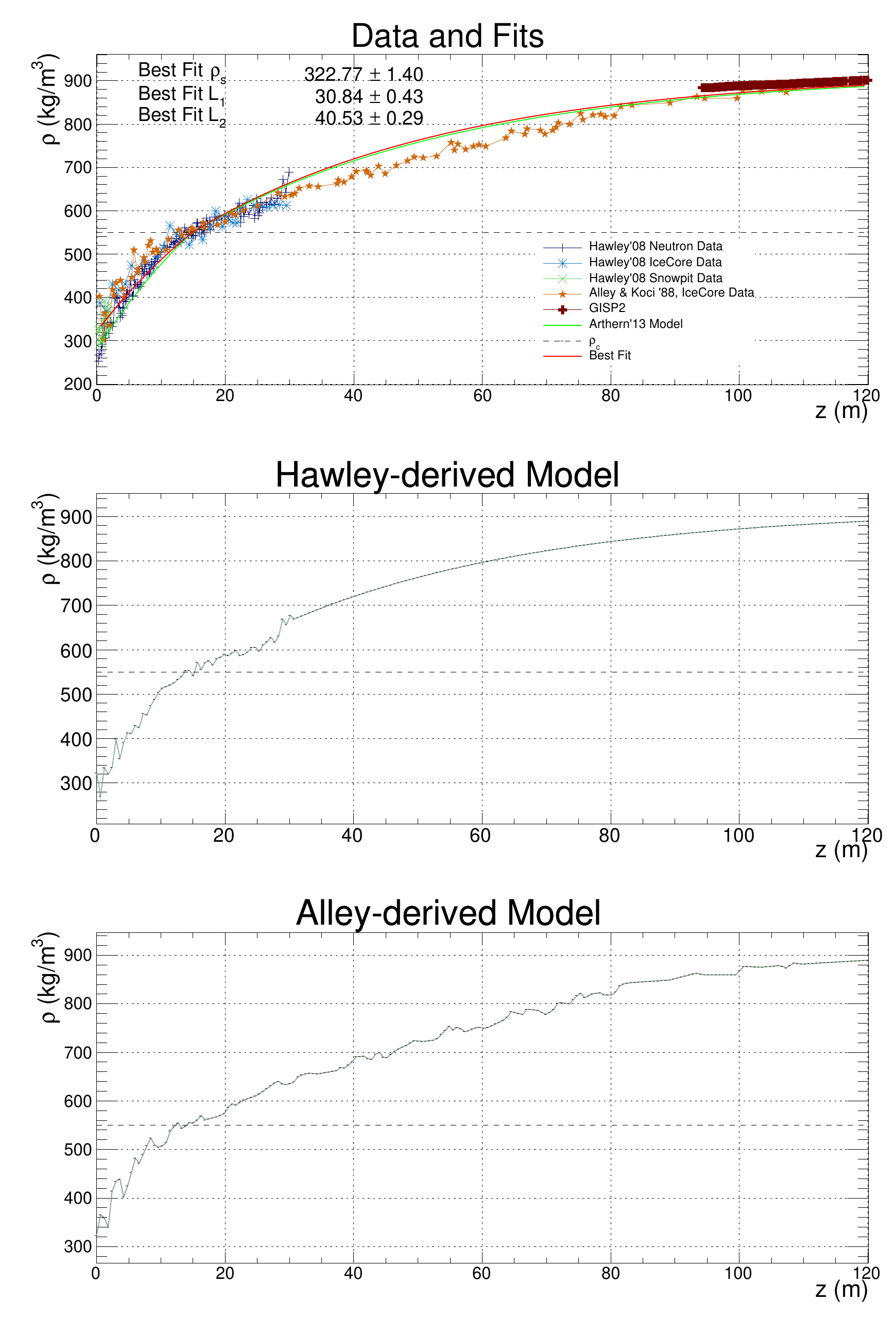}
    \end{center}
    \caption{Top: Various firn density data sets recorded at or near Summit Station, using a variety of 
      techniques~\cite{hawley, alley, gisp2}, and a model based on radio echo data at Summit Station~\cite{arth}.  We
      fit a double exponential to all of the data shown, motivated by~\cite{herron} and shown with the red line.
      From this, we choose to use two firn models for our analysis.  Middle: The first model 
      uses the Hawley neutron density probe data shallower
      than 30~m and the best-fit exponential deeper than 30~m~\cite{hawley}.  Bottom: The second model uses the Alley ice core
      data to 100~m, and the best-fit exponential at larger depths~\cite{alley}. The parameters of the best-fit double
      exponential are shown, where $\rho_s$ is the density at the surface in kg/m$^3$, and $L_1$ and $L_2$ are the length
      scales in meters of each portion of the exponential fit. 
      }
    \label{fig:firnStuff}
  \end{figure}

Recently, observations of horizontally-propagating waves between transmitters
deployed in the ice in shadowed regions relative to the receivers indicate that
other propagation modes exist~\cite{barwick}.  Modes that include ``surface,''
``lateral,'' and ``leaky'' waves are well-known solutions to Maxwell's Equations at dielectric
interfaces (see e.g.~\cite{lateral} for a pedagogical
review).  Previously, surface waves have been discussed as a possible mechanism for neutrino 
detection~\cite{ralston}, but
previous measurements show no evidence for significant power in a
surface-wave mode~\cite{taylorDome,Alvarez}.  The recently-observed propagation
consistent with a horizontally-propagating mode at arbitrary depth~\cite{barwick} has
implications for determining the optimal geometry of neutrino detectors if the
coupling to the horizontally-propagating modes is large.  If the power
contained in these modes is small compared to the ray-bending mode, then
signals in this region would not contribute significantly to the effective
volume of in-ice neutrino detectors.

We have made measurements of radio propagation through the firn at Summit
Station, Greenland in June 2013 using a fast, impulsive, high-voltage
transmitter placed a few feet below the surface and a receiver lowered down a
borehole, up to 1050~ft. away and 600~ft. deep.  At the largest depths, the
receiver is well below the firn layer at Summit, which is measured to be
$\sim~100$~m~\cite{arth}.  We compare these measurements to models of
electromagnetic wave propagation through the firn to constrain possible radio
propagation modes. 

In Section~\ref{sec:modeling}, we discuss firn density models at Summit Station, 
a finite-difference time-domain (FDTD) simulation, and a ray-tracing
algorithm that explores the bulk ray-bending propagation mode as well 
as surface-wave and arbitrary-depth horizontal propagation modes.  In Section~\ref{sec:measurements}, we present
measurements of radio propagation through the firn at Summit Station
in the $\sim100$ MHz range.  We then discuss the implications for 
radio detectors for high-energy neutrinos interacting in glacial ice in Section~\ref{sec:discussion}.

\section{Modeling of Radio Propagation in the Firn at Summit Station, Greenland}
\label{sec:modeling}
\subsection{Firn Index of Refraction Model}
\label{sec:firnModel}
To compare with measurements, we construct two different firn models, based on a variety of measurements of
the density profile,~$\rho(z)$,~at and near Summit Station~\cite{hawley,alley,gisp2,arth}
to determine the index of refraction as a function of depth in meters ($z$) for the firn layer.
The commonly-used Herron-Langway parameterization of firn models has two density regimes, each of which is fit by 
an exponential subtracted from an offset~\cite{herron}.  
The break point between the two exponential fits occurs at a depth corresponding
to the critical density, $\rho_c$=550 kg/m$^3$, above which the onset of jamming of snow grains leads to slower 
compactification~\cite{herron}.
The data that we use to construct our models is shown in Figure~\ref{fig:firnStuff}.  We perform 
a double exponential fit to all of the data, shown with the red line in Figure~\ref{fig:firnStuff}.  The fit parameters
are shown in Figure~\ref{fig:firnStuff} as well, and using these parameters, we find a critical depth of 14.9~m and
that $\rho(z)$ is given by:

\begin{equation}
\begin{split}
\rho(z)=0.917 - 0.594e^{-z/30.8}   \qquad z\le 14.9 \\
\rho(z)=0.917 - 0.367e^{-(z-14.9)/40.5} \qquad  z > 14.9.
\end{split}
\end{equation}
\begin{figure*}[]
      \begin{center}
        \includegraphics[width=\textwidth]{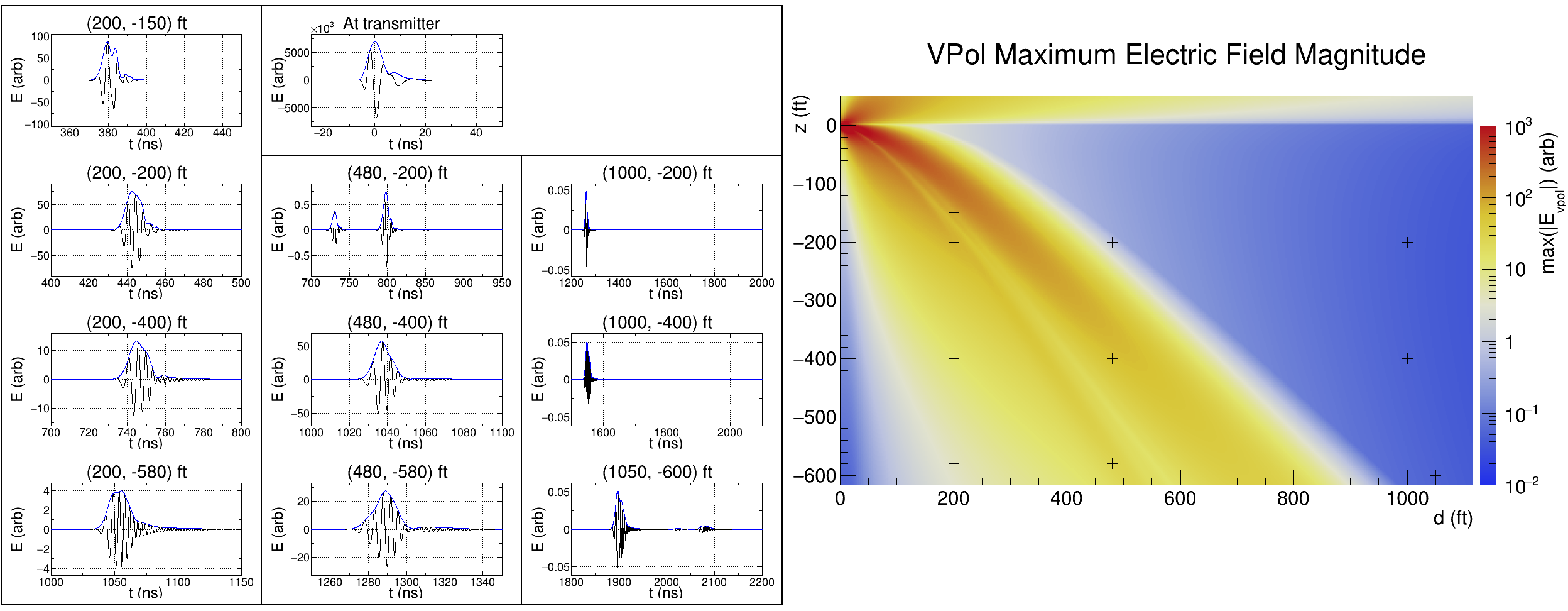}
    \end{center}
    \caption{The result from the FDTD simulation using a best-fit of available data to a Herron-Langway firn model, described in Section~\ref{sec:firnModel}. Although the model is continuous, it is not differentiable at the critical density, producing an additional set of reflections. In the simulation,
      a dipole transmitter is placed 3~ft. below
      the surface of the snow, from which a band-limited impulse between 90-250~MHz propagates through the firn.  
      The color map shows the maximum vertically-polarized 
      electric field reached over the course of the simulation at each point.
      Each cross on the color map indicates a geometry where we placed a receiver at Summit Station relative to the transmitter location, which is set at (0, -3 ft.) here (see Section~\ref{sec:setup} for further discussion).
      Resulting simulated waveforms (electric field
      as a function of time) are shown for the location of each cross and for the location of the transmitter.  
      The relative amplitudes of the waveforms can be compared.
      }
    \label{fig:fdtdFit}
  \end{figure*}
\begin{figure*}[]
      \begin{center}
        \includegraphics[width=\textwidth]{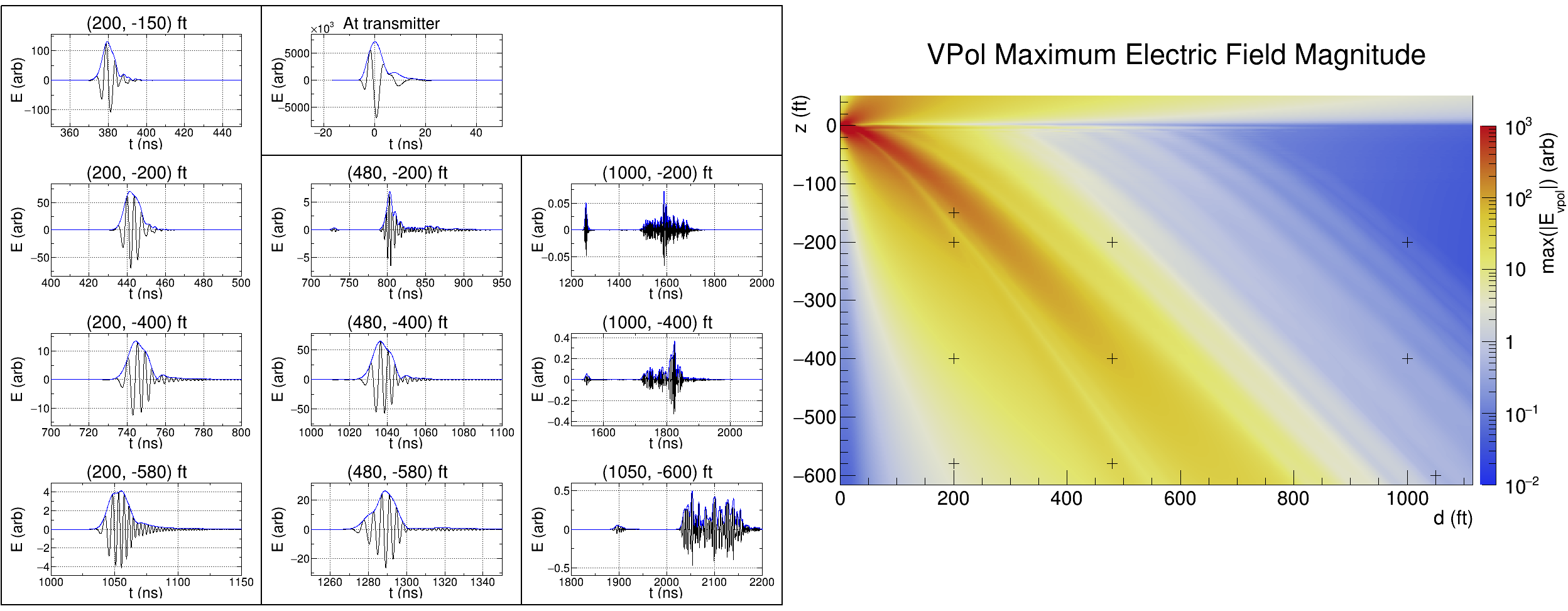}
    \end{center}
    \caption{The result from the FDTD simulation using a firn model based on neutron scattering data from 
      Hawley~\cite{hawley} and a
      global exponential fit to all data sets beyond 30~m, described in Section~\ref{sec:firnModel}.  In the simulation,
      a dipole transmitter is placed 3~ft. below
      the surface of the snow, from which a band-limited impulse between 90-250~MHz propagates through the firn.  
      The color map shows the maximum vertically-polarized
      electric field reached over the course of the simulation at each point.
      Each cross on the color map indicates a geometry where we placed a receiver at Summit Station. Resulting 
      Each cross on the color map indicates a geometry where we placed a receiver at Summit Station relative to the transmitter location, which is set at (0, -3 ft.) here (see Section~\ref{sec:setup} for further discussion).
      Resulting simulated waveforms (electric field
      as a function of time) are shown for the location of each cross and for the location of the transmitter.  
      The relative amplitudes of the waveforms can be compared.
      }
    \label{fig:fdtdHawley}
  \end{figure*}

Horizontal propagation modes require
perturbations from a smoothly-varying density gradient, motivating the 
use of models that include the real density fluctuations 
observed at Summit Station.  We therefore 
construct two models that use different linearly-interpolated
data sets over their entire range, and the best-fit exponential elsewhere.
The first model consists of the neutron scattering data from Hawley~\cite{hawley} at depths smaller
than 30~m and the best-global-fit exponential beyond 30~m.  The second model uses measured densities from
the Alley ice core data 
up to 100~m, since it is the only data set that spans such a large range of depths, and then the best-global-fit
exponential beyond 100~m.  By a depth of $\sim100$~m, the firn has transitioned to glacial ice.  These 
firn models are not perfect -- the density profile changes each year due to snow accumulation and compaction
and varies from site to site, the firn layers may be under sampled in the data,
making both the neutron scattering and ice core data imperfect for our purposes.

For radio propagation in glacial ice, the dielectric constant ($\epsilon$)
is related to the density (in g/cm$^3$) of the ice~\cite{kovacs} by:

\begin{equation}
\epsilon ' = (1+0.845 \rho)^2,
\end{equation}
allowing us to calculate the index of refraction $n=\sqrt{\epsilon'}$ 
as a function of depth for the firn.

\subsection{Finite-Difference Time-Domain (FDTD) Simulation}
\label{sec:sims}

Previous studies of radio propagation in ice have used geometric optics, which
is valid in regimes where the wavelength is much smaller than the
feature sizes. Layers of ice near the surface contain features of sizes
comparable to wavelengths (including the ice-air interface), so we have implemented the
FDTD method~\cite{fdtd}, which numerically
solves the wave equation on a time-space lattice. This powerful but
computationally expensive time-domain treatment is particularly appropriate for
propagation of the impulsive broadband emission expected to be
generated by neutrinos. Our simulation software interfaces with the
FDTD solver \texttt{meep}~\cite{meep}~\footnote{Our simulation code is available
at \url{https://github.com/cozzyd/iceprop}}. 

We use the cylindrical symmetry of a vertical dipole antenna to reduce the
computation requirements. The transmitter
is at horizontal distance $d=0$ and the ice surface is at $z=0$. $n(z)$ varies
according to the chosen ice density model below the ice surface, and $n$=1
above the surface. We do not include attenuation or absorption in the ice in the simulation.
Perfectly-matched layers (PMLs), which are absorptive across a range of
frequencies, are placed at the edges of the computational domain to
simulate propagation outside of the domain (otherwise waves would reflect off the
edges of the computational domain). %

We simulate a dipole transmitter at -3~ft. in a
domain with $z \in[190, 80]$ m and $d\in[0,360]$ m, with a 20~m PML located
outside the volume.  To keep the computational requirements
modest, we use a grid resolution of 5~cm. At this
resolution, numerical dispersion limits the maximum frequencies that may be
reliably simulated to 300~MHz. We simulate a vertical dipole electric-field impulse with
with a real component corresponding to the impulse response of a 90-250 MHz fourth-order
digital Butterworth bandpass filter to roughly match the frequency content of
our measurements, which are reported in Section~\ref{sec:measurements}. 

\begin{figure*}[]
      \begin{center}
        \includegraphics[width=\textwidth]{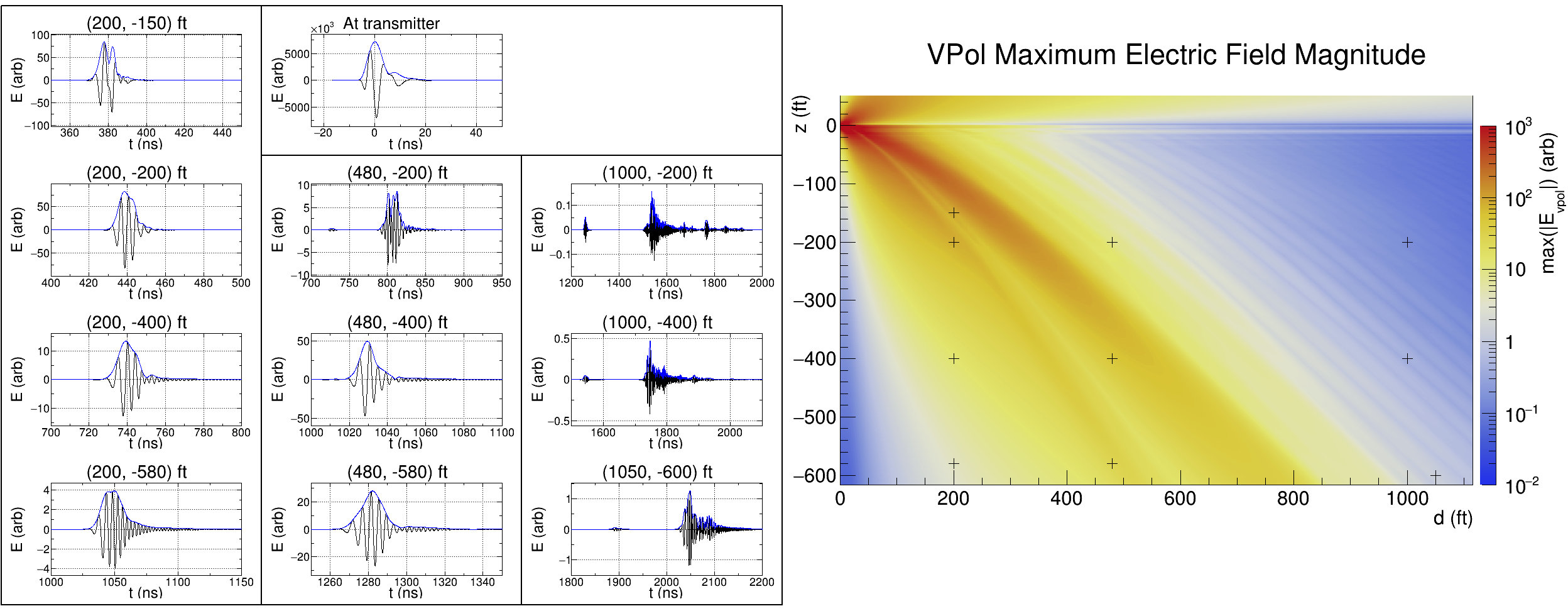}
    \end{center}
    \caption{The result from the FDTD simulation using a firn model based on ice core data from Alley~\cite{alley} and a
      global exponential fit to all data sets beyond 100~m, described in Section~\ref{sec:firnModel}.  In the simulation,
      a dipole transmitter is placed 3~ft. below
      the surface of the snow, from which a band-limited impulse between 90-250~MHz propagates through the firn.  
      The color map shows the maximum vertically-polarized
      electric field reached over the course of the simulation at each point.
      Each cross on the color map indicates a geometry where we placed a receiver at Summit Station relative to the transmitter location, which is set at (0, -3 ft.) here (see Section~\ref{sec:setup} for further discussion).
      Resulting simulated waveforms (electric field
      as a function of time) are shown for the location of each cross and for the location of the transmitter.  
      The relative amplitudes of the waveforms can be compared.
      }
    \label{fig:fdtdAlley}
  \end{figure*}

We perform simulations with the ice models described in Section~\ref{sec:firnModel}
above and record the maximum electric field magnitude achieved on a 0.5 m
grid. We also record complete time domain information at the transmitter and
at positions corresponding to our measurements. The results from the
simulations are shown in Figures~\ref{fig:fdtdFit},~\ref{fig:fdtdHawley}, and~\ref{fig:fdtdAlley}.
We note that beam pattern suppression from the dipole transmitter dominates
at steep angles.

\begin{figure*}[]
      \begin{center}
        \includegraphics[width=0.45\textwidth]{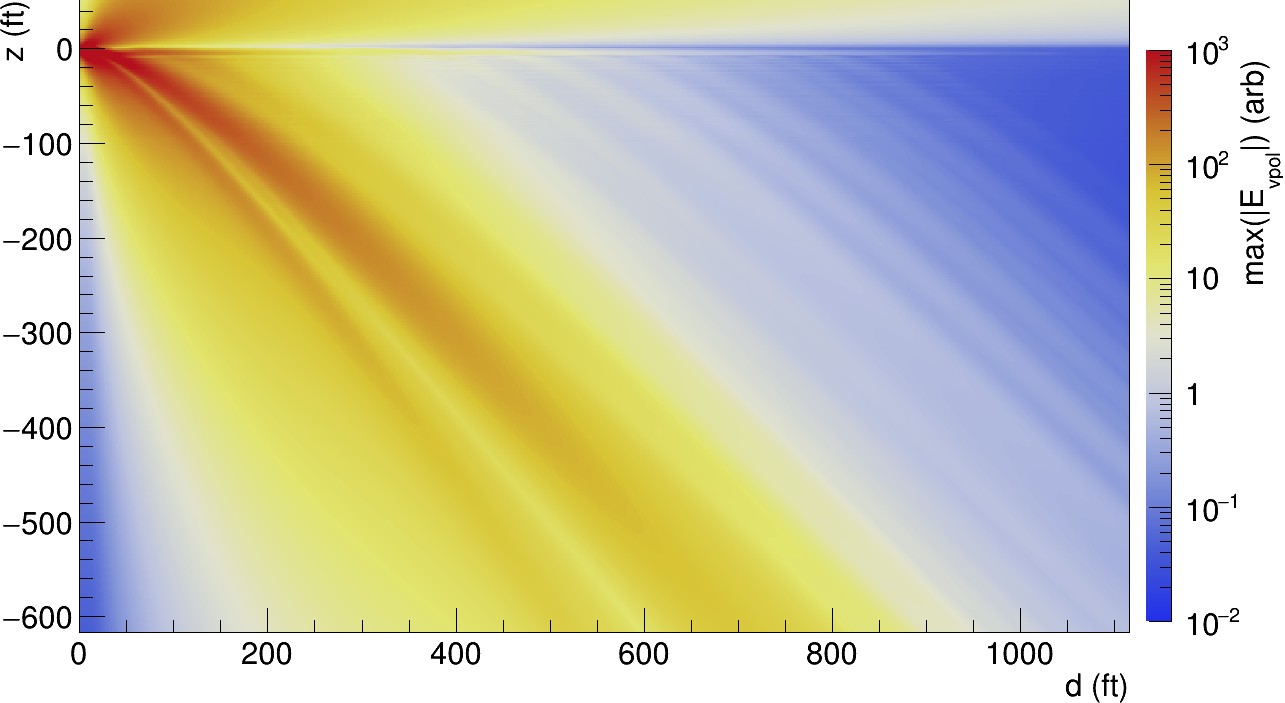}
        \hspace{0.05\textwidth}
        \includegraphics[width=0.45\textwidth]{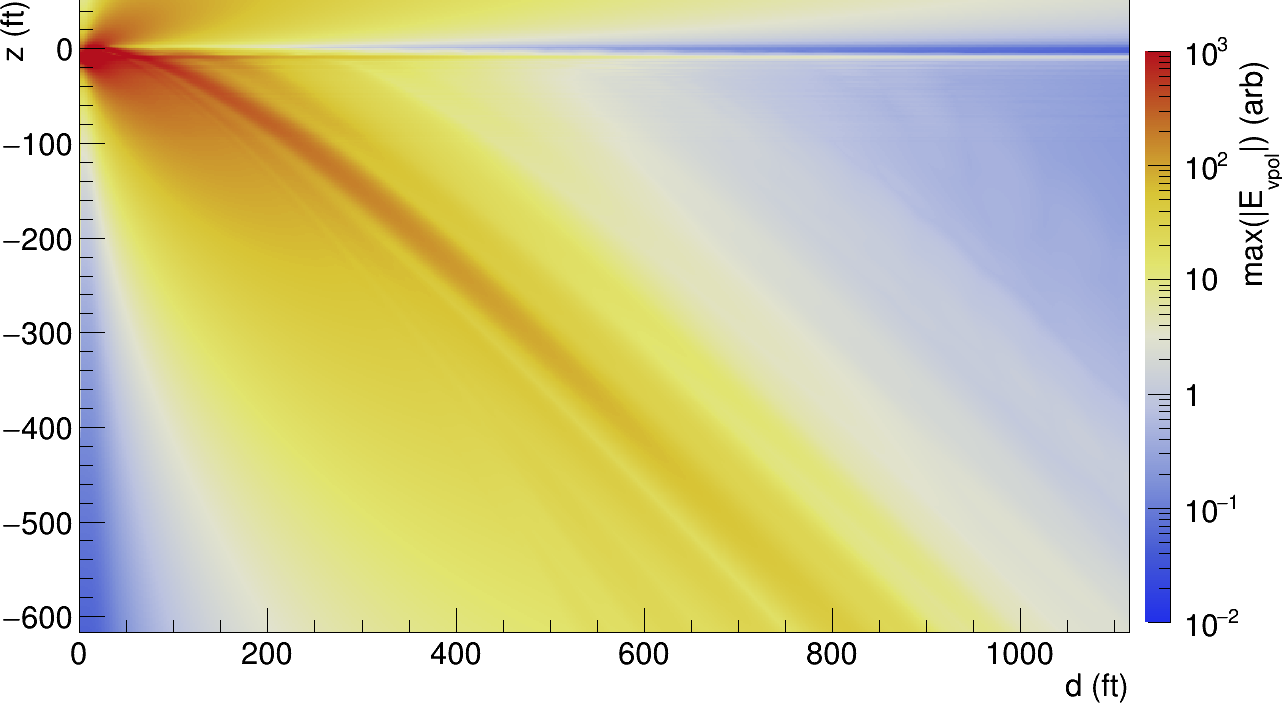}      \\
        \includegraphics[width=0.45\textwidth]{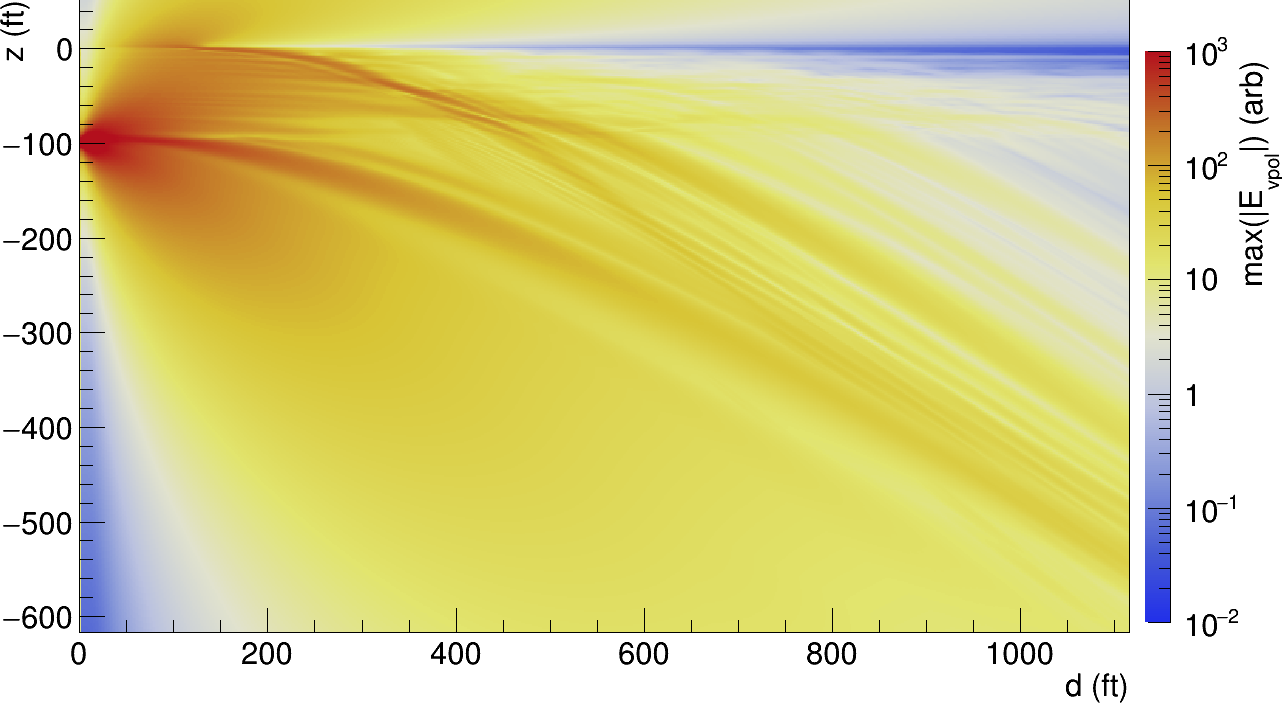}
        \hspace{0.05\textwidth}
        \includegraphics[width=0.45\textwidth]{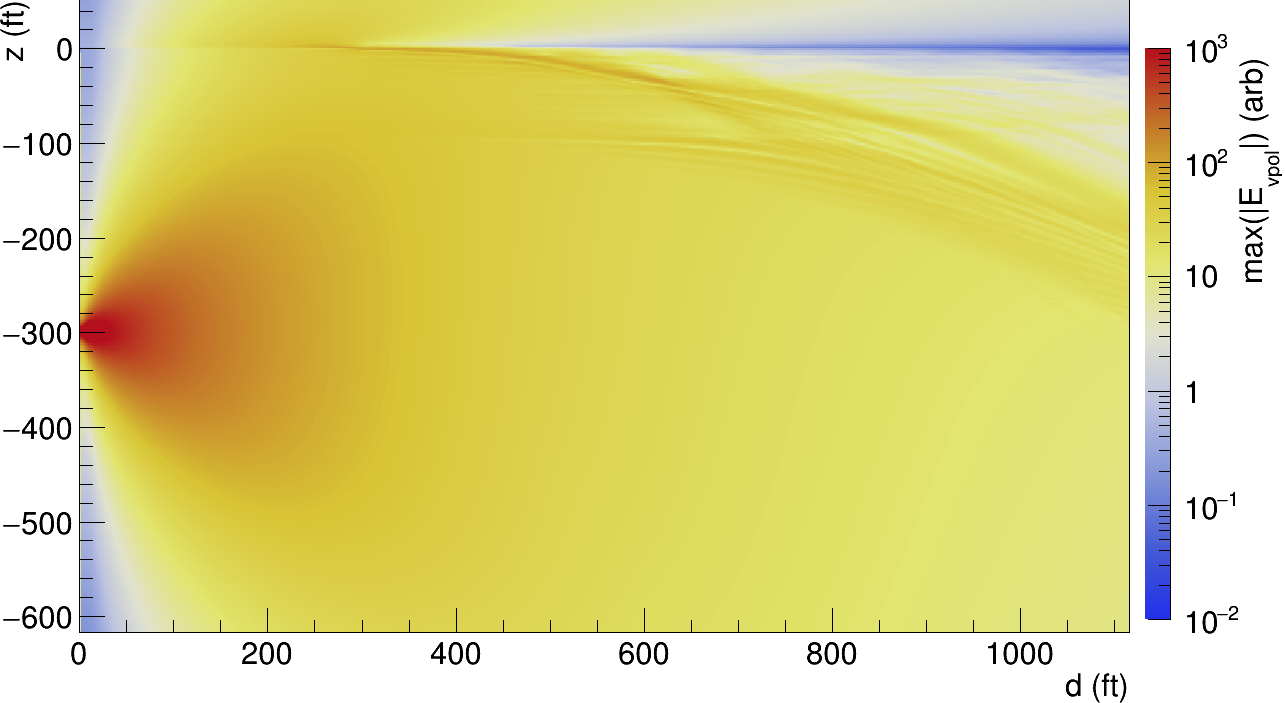}
      \end{center}

    \caption{Variations on the FDTD simulation in Figure~\ref{fig:fdtdHawley}. 
      Upper Left: The signal is 200-300 MHz instead of 90-250 MHz. 
      Upper Right: The transmitter is placed 10~ft. below the surface instead of 3 ft.
      below.  Lower Left: The transmitter is placed 100~ft. below the surface. Lower Right: The transmitter is placed
      300~ft. below the surface.      
      The firn model, based on neutron scattering data from Hawley~\cite{hawley} is kept the same.  The color map shows the 
      maximum vertically-polarized electric
      field magnitude reached over the course of the simulation at each point.}
    \label{fig:fdtdVariations}
\end{figure*}

To investigate the effect of the depth of the transmitter and the frequency content
of the signal, we vary these parameters in the simulation, and show
the results in Figure~\ref{fig:fdtdVariations}.  As the transmitter is lowered,
the refractive ray-bending becomes less pronounced, and reflections off of the ice/air interface
serve to further fill in the shadowed region.  The simulations at all depths also show power beyond the 
predicted shadowed region (even after accounting for reflections off of the surface), which
supports the idea that waves propagate along density perturbations in the firn.  The amplitude
of the signals seen in simulations in this region is significantly smaller than in the non-shadowed region
at all depths.

\begin{figure}[]
      \begin{center}
        \includegraphics[width=8.5cm]{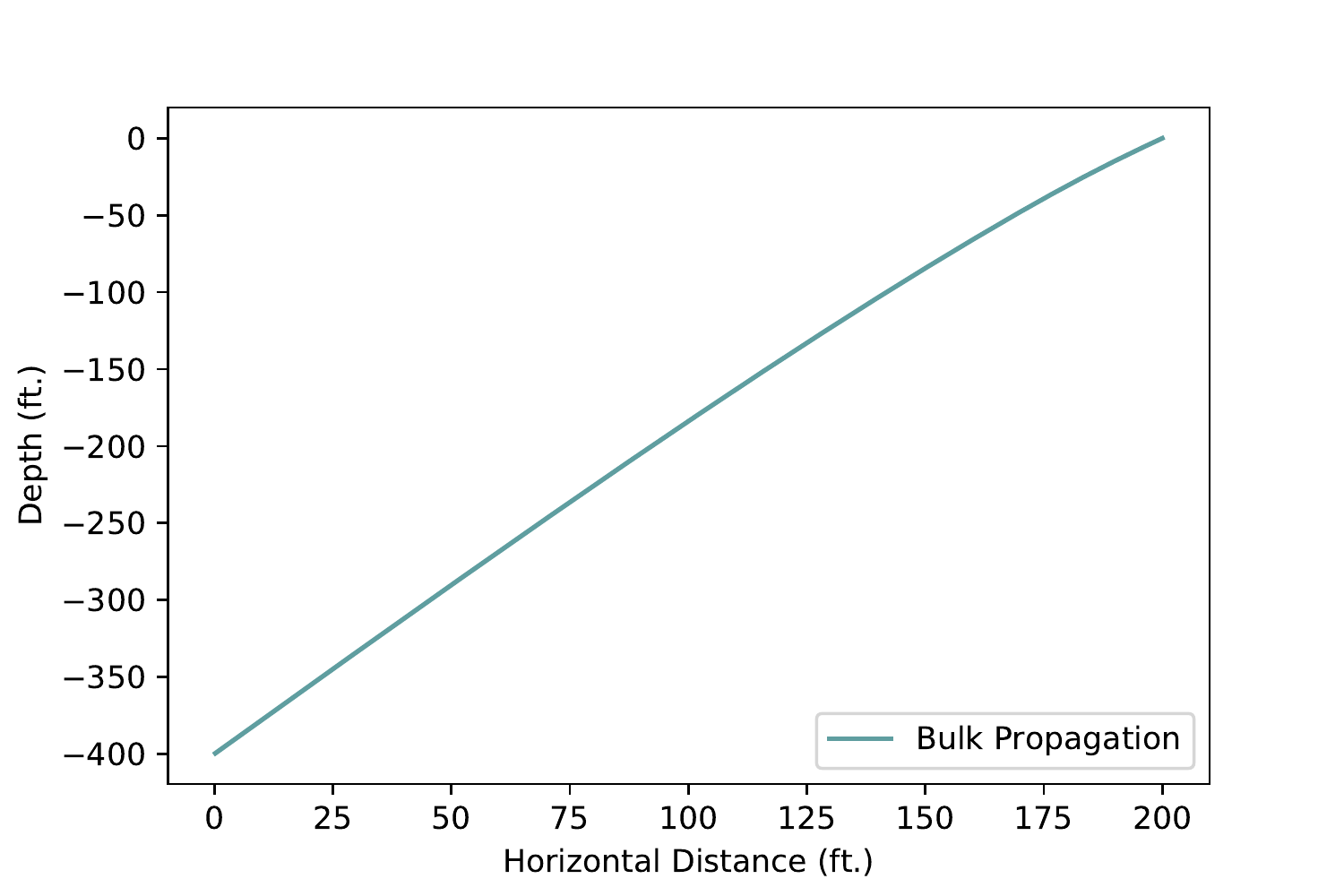}
        \includegraphics[width=8.5cm]{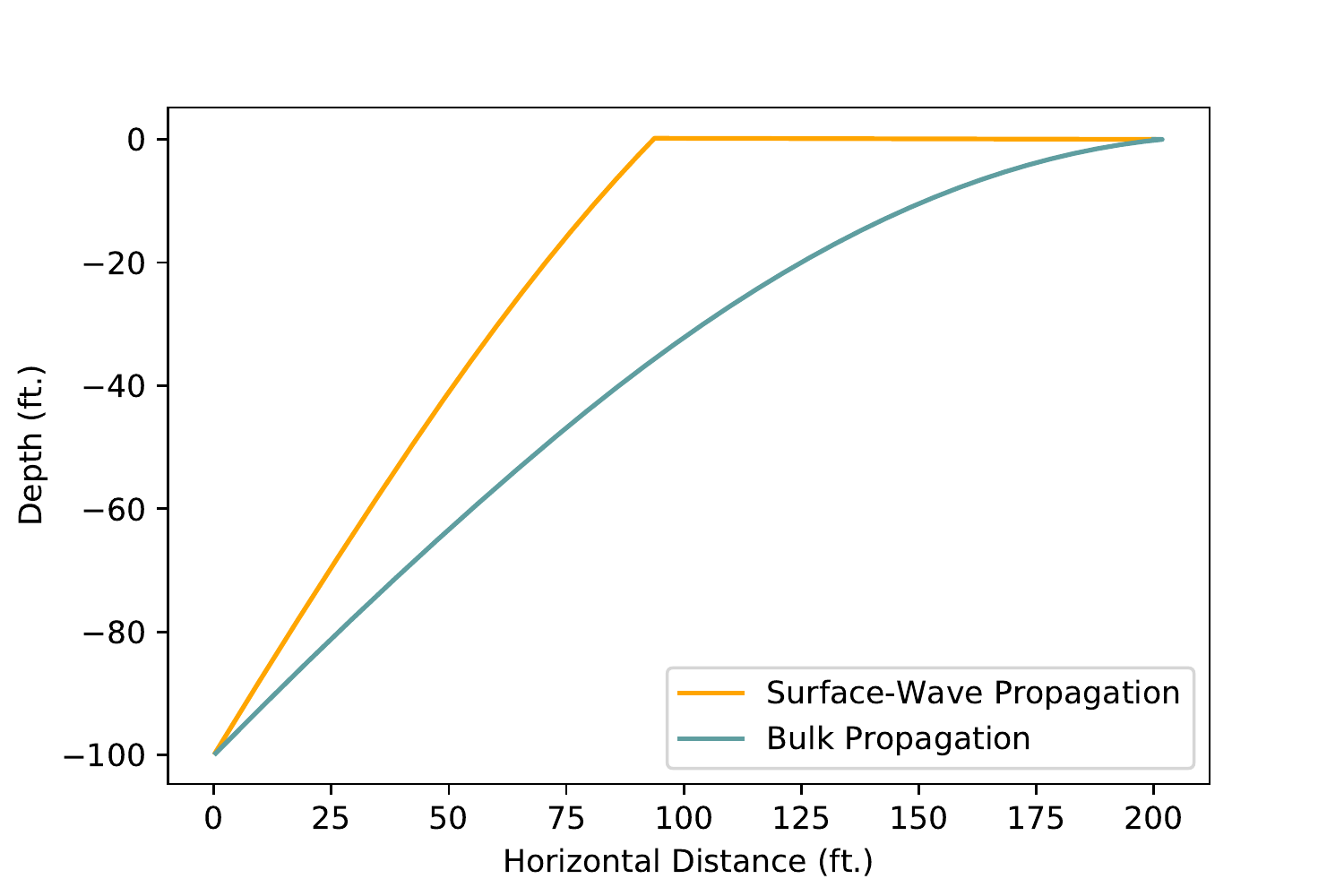}
        \includegraphics[width=8.5cm]{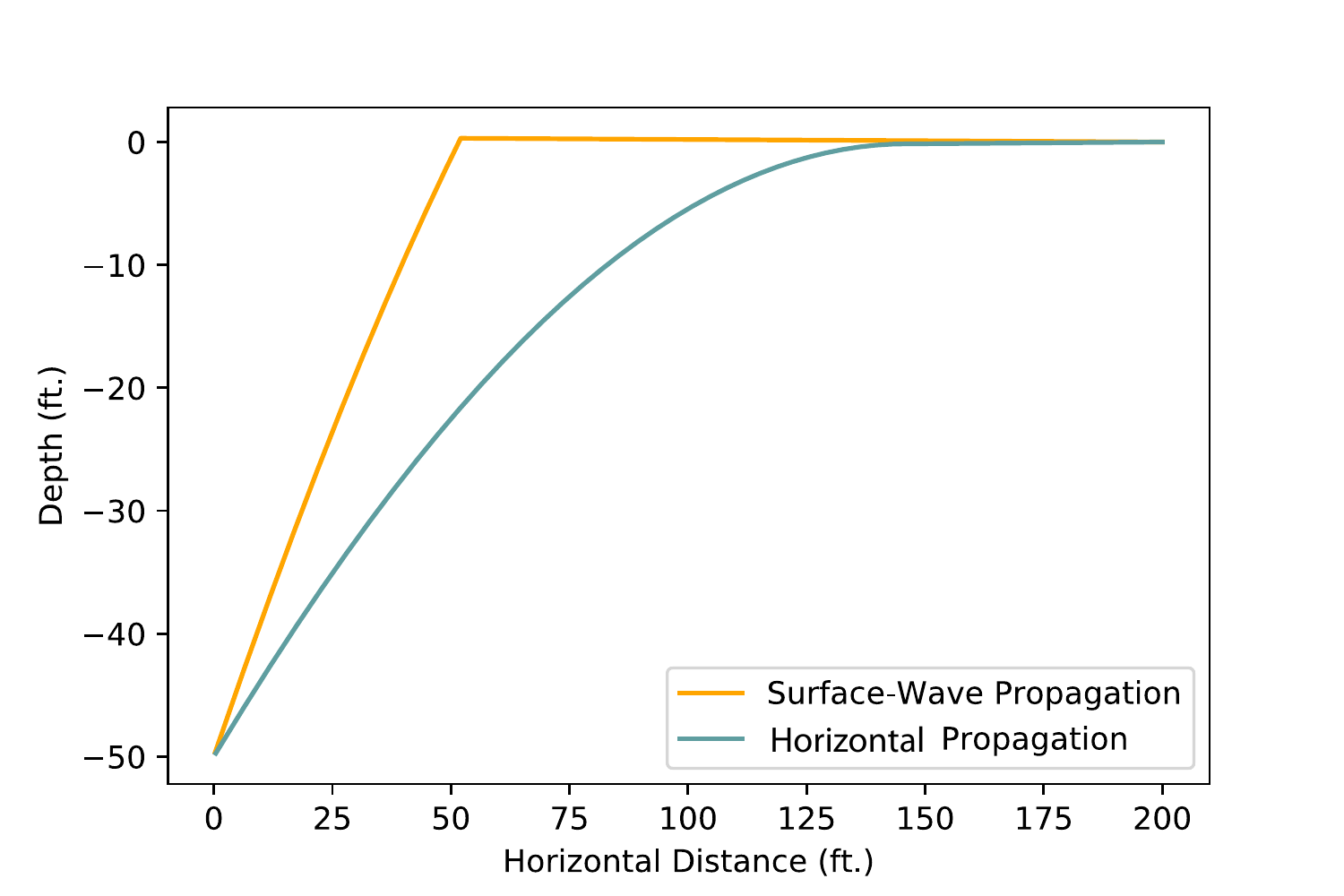}
    \end{center}
    \caption{Example ray-tracing solutions given a transmitter and receiver depth.  
      In each case shown here, the transmitter is just below the surface.
      Top: A geometry where only the 
      bulk propagation (refractive ray-bending) mode yields
      a solution (Receiver Depth 400 ft., Transmitter Horizontal Distance 200~ft.).  
      Middle: A geometry where the bulk propagation mode yields
      a solution and there is an additional surface-wave solution~\cite{ralston} up to the TIR angle 
      (Receiver Depth 100 ft., Transmitter Horizontal Distance 200~ft.).  Bottom:  
      A geometry where the bulk propagation mode yields no 
      solution, but a horizontally-propagating solution~\cite{barwick} is found
      (Receiver Depth 50 ft., Transmitter Horizontal Distance 200~ft.).  In these geometries, there
      is always also an additional surface-wave solution up to the TIR angle~\cite{ralston}.
      }
    \label{fig:rayTrace}
  \end{figure}

The simulations indicate that although the signal is strongest in the
non-shadowed zone, a signal is still present in all regions.
For the non-smooth firn models, the region corresponding to the shadow of
the transmitter contains two timescales of propagation.  The earlier-arriving
waves are present even with a smooth density profile, and are
identified with ``surface''  or ``lateral" waves.   The second kind of
propagation produces a longer train of signals and does not appear in a smooth
density model, so it is likely from reflections between layers or
channeling along peaks or troughs in the density profile.  The details of the
second signal depend strongly on the firn model, transmitter depth,
and frequency content of the signal. Qualitatively, the relative amplitude of
the first and second signal changes as a function of depth, with the first
signal growing stronger with respect to the second at larger depths.

On top of the effect of the beam pattern of the simulated transmitter, the
simulation predicts additional significant spatial variations in maximum
electric field having to do with interference between different paths, even in
the non-shadowed region, visible in the colormaps in
Figures~\ref{fig:fdtdHawley} and~\ref{fig:fdtdAlley}.  The presence of these
amplitude variations makes amplitude predictions dependent on the ice
model and frequency content of the signal. 

\subsection{Ray Tracing}
\label{sec:models}

We perform additional ray-tracing
studies to investigate if it is possible to model in a simple way the three different
modes suggested by the FDTD simulations in Section~\ref{sec:sims}.  We place a
transmitter near the surface and a receiver at a chosen depth, $z$, and
horizontal distance, $d$, from the transmitter, and find the ray that connects
the transmitter and receiver location for each of three modes.  The first mode is
the bulk refractive propagation mode, corresponding to standard geometric optics,
which predicts ray-bending in the firn and therefore a shadowed region.  The
second is a surface-wave propagation mode, discussed in~\cite{ralston}, where
for rays that hit the ice/air interface at the total internal reflection (TIR)
angle ($48^\circ$) or steeper, a fraction of the power will couple to a surface
wave.  Here we model this as propagating along the surface and losing electric field strength
as $1/\sqrt{r}$ due to being confined to two dimensions.  The third is an
arbitrary-depth horizontal propagation mode, where for rays that become
horizontal due to refractive ray-bending, a fraction of the power couples to a
horizontal mode if there is a specific class of perturbation from a smooth
density gradient at that depth, as may occur in firn layers due to annual snow
deposits~\cite{barwick}.  We allow for two-dimensional ($1/\sqrt{r}$) 
or three-dimensional ($1/r$) propagation.  We call these modes ``bulk'', ``surface'', and
``horizontal'' propagation, respectively.

For a transmitter placed just below the surface, one or two of the three
ray-tracing propagation modes will find a 
valid solution for an arbitrarily-chosen $z$ and $d$, as shown in 
Figure~\ref{fig:rayTrace}. In the non-shadowed region (top and middle panels), there 
can also be a set of surface-wave solutions (middle panel).  In the shadowed
region (bottom panel), there is always both a horizontal and surface-wave solution for a transmitter
placed near the surface.

\section{Firn Propagation Measurements at Summit Station, Greenland}
\label{sec:measurements}
\subsection{Experimental Setup}
\label{sec:setup} 
\begin{figure}[]
      \begin{center}
        \includegraphics[width=9cm]{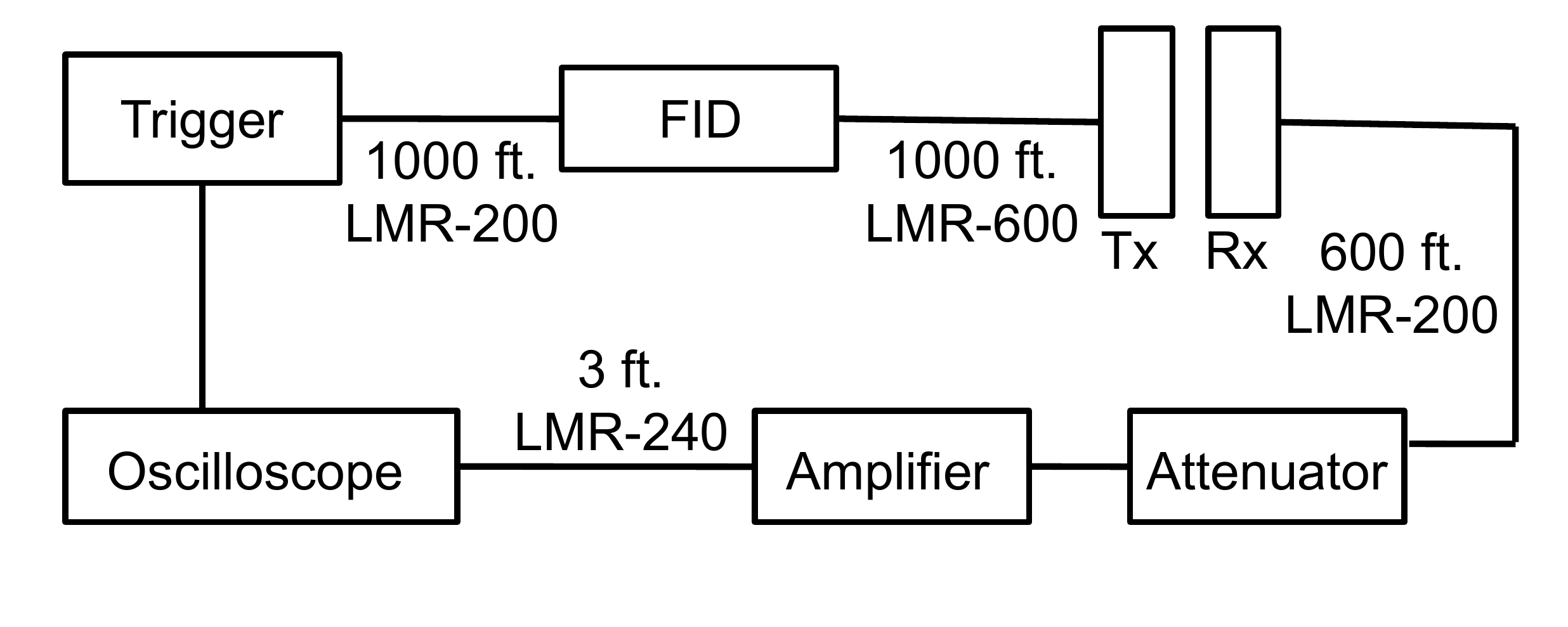}
    \end{center}
    \caption{A schematic of the electronics layout for our experimental setup.  A 6~kV FID Technologies pulser triggered
      by an external TTL signal was 
      used to send signals to the transmitting low-gain antenna.  
      After the signal was received
      by the receiving antenna, it was attenuated (in some geometries) and then amplified using a low-noise Miteq
      amplifier and read out using an oscilloscope, which was triggered using the same TTL signal that triggered
      the FID Technologies pulser.
     }
    \label{fig:airSchematic}
  \end{figure}

A schematic of our experimental design is shown in Figure~\ref{fig:airSchematic}.  We used
a 6~kV FID Technologies pulser triggered by an external TTL signal to generate impulsive radio signals with 
power in the 100~MHz -- 1~GHz range.  We transmitted the signal from a low-gain fat dipole antenna, built for the 
RICE experiment~\cite{rice} at the South Pole.  An identical
receiving antenna was placed a distance away, and the received signal was first sent through
an attenuator (as needed, depending on the physical configuration of the transmitter and receiver), 
followed by a Miteq AFS3-00200120-10-1P-4-L amplifier with 50~dB of gain.  We read the signals out using a Tektronix
MSO5204B oscilloscope, which was triggered on a different channel 
using the same signal generator that triggered the high voltage FID Technologies 
pulser.  The oscilloscope recorded waveforms with 10,000 averages so that the noise 
level in recorded waveforms is small.  We used Times Microwave LMR-600, LMR-240, 
and LMR-200, and cable lengths are given in Figure~\ref{fig:airSchematic}.  

\begin{figure}[]
      \begin{center}
        \includegraphics[width=8.5cm]{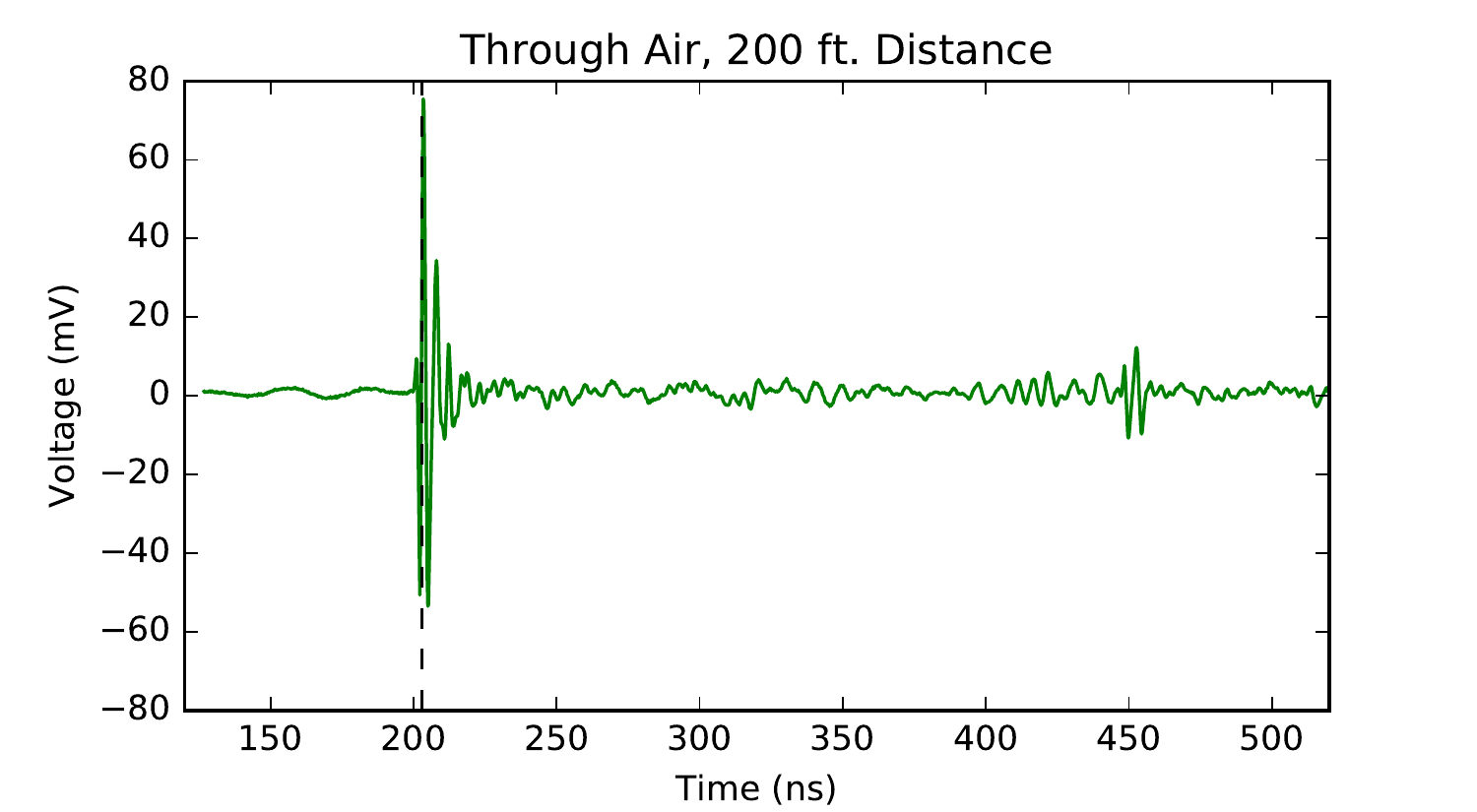}
    \end{center}
    \caption{The received signal, using the experimental setup shown in Figure~\ref{fig:airSchematic}, with the transmitting
      and receiving antennas separated by 200~ft. in the air.  For this measurement, the 
      antennas were placed $\sim4$~ft. above the surface of the snow on empty wooden crates.  The time shown
      is the absolute propagation time of the signal after removing the system delay.  The dashed line
      shows the predicted time of flight.  The second pulse seen at 245~ns
      after the initial pulse is a known reflection in the system corresponding the out-and-back time along one section 
      of LMR-600 cable, and appears in all data. Because we did not record the overall attenuation setting
      for this measurement, the amplitude information is not comparable to the downhole measurements.
    } 
    \label{fig:airWaveform}
  \end{figure}
\begin{figure}[]
      \begin{center}
        \includegraphics[width=7cm]{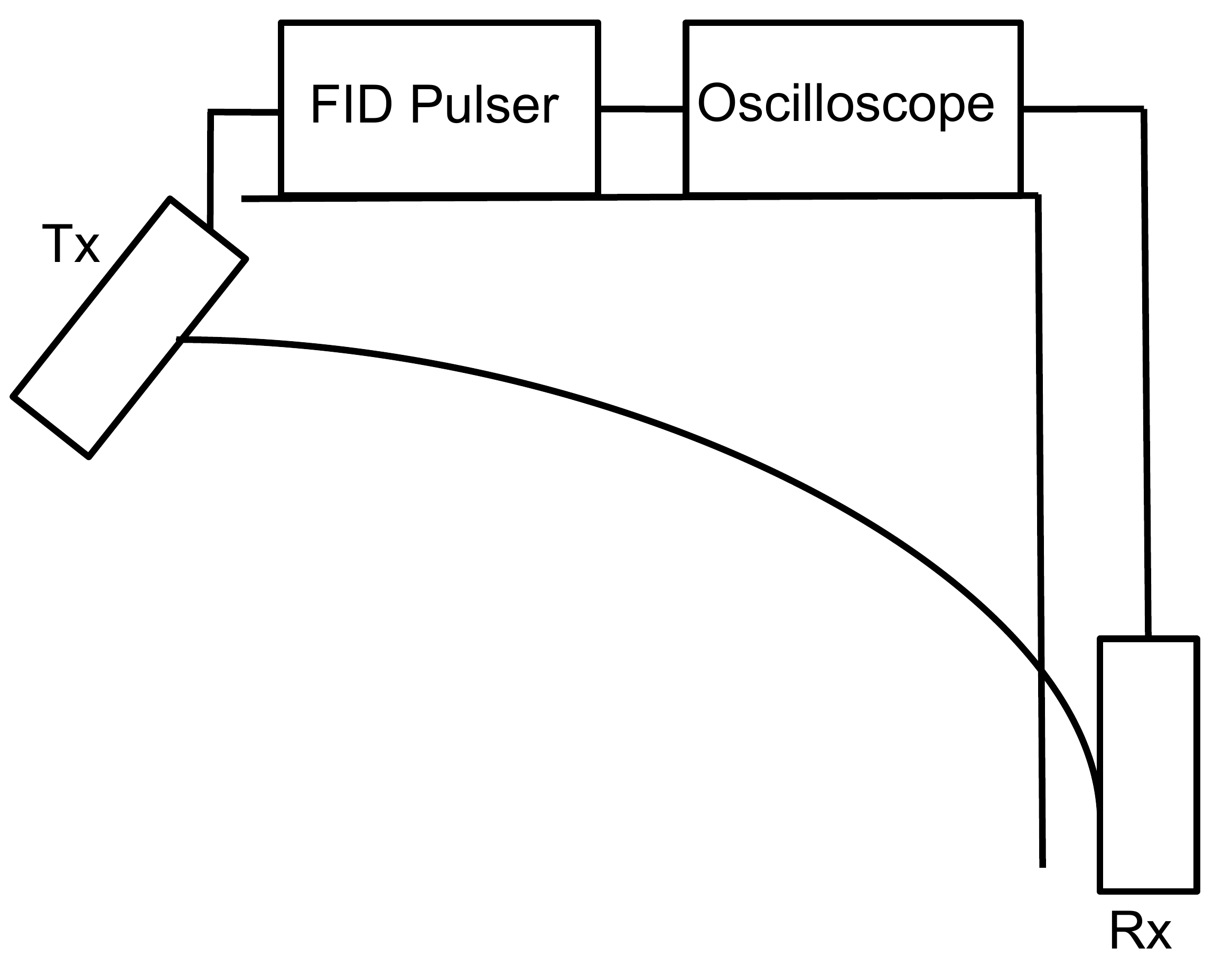}
    \end{center}
    \caption{A simplified schematic of the experimental setup, as deployed at Summit Station, Greenland.  The receiving
      antenna was placed down a borehole that was filled with drilling fluid to $\sim100$~ft. depth, and the feed 
      of the transmitting antenna was placed $\sim~3$~ft. below the surface using an auger drill. 
      The front-end electronics and triggering scheme are the same as in the in-air measurements 
      shown in Figure~\ref{fig:airSchematic}. 
    \label{fig:downholeSchematic}}
  \end{figure}

To determine the absolute delay of our system, we 
first made a measurement with the transmitter and receiver separated by 200~ft. horizontally 
in the air.  The antennas were placed on top of wooden crates, $\sim4$~ft. above the surface of the snow.
The received in-air waveform is shown in Figure~\ref{fig:airWaveform}.  The in-air signals show a system 
delay of 2395~ns, which we then subtract from each subsequent measurement of time of flight with the receivers 
placed in the ice. We did not record the overall attenuation setting for the in-air measurement relative 
to the downhole measurements discussed below, so the amplitude of the in-air pulse is not comparable to 
the downhole measurements.  The in-air pulse measured with our system has a pass band from 190--270~MHz, measured 
at its -3~dB points and defined by the
antenna and amplifier response and the loss in the cables as a function of frequency.  
When the antennas are embedded in the ice, we expect the frequency 
response of the antennas to go down.

There is a reflection in our system, obvious in the data in Figure~\ref{fig:airWaveform}, that arrives 245~ns
after the initial pulse.  The relative timing of the reflection did not change as we changed the distance between
the antennas or moved the antennas closer or farther from the surface.  We surmise that this is from
a reflection in the system, as it is present in all data 
and corresponds to the out-and-back propagation delay from 100~ft. of LMR-600 cable, 
segments of which are present in the system.  Other than this
reflection in the system, the in-air measurements show a clean impulse.

We took data using our system with a variety of transmitter and receiver geometries.  The layout
is shown in Figure~\ref{fig:downholeSchematic}.  We lowered the receiver down 
the DISC borehole at Summit Station, which has a plastic casing along the upper $\sim100$~ft. and 
is filled with non-conductive
drilling fluid below $\sim100$~ft.  We used an auger drill to make shallow holes at the surface to place
the transmitting antenna into, so that the antenna feed was $\sim3$~ft. below the surface of the snow to 
ensure good coupling to the snow.  At each transmitter location, we drilled these 
shallow holes at a variety of angles with respect to the vertical and looked for the angle at 
which the signal strength was strongest to ensure that the antenna was broadside to the path of propagation.  We 
could not perform a similar procedure for the receiver down the borehole, so we calculate a correction to the amplitude
due to the beam pattern of the dipole antenna down the hole based on the incidence angles predicted by 
our ray tracer and a $\sin\theta$ antenna beam pattern in electric field to use as an uncertainty in the
calculations in Section~\ref{sec:coupling}. 

When the receiving antenna passed
below the level of the drilling fluid in the borehole, we saw an increase in signal strength, 
which indicates improved coupling to the ice.  This effect was repeatable and clearly related to entering 
and exiting the fluid. We therefore only report on measurements at depths below 100~ft., since there is a discontinuity
in amplitudes at a depth of 100~ft.

\begin{table*}
\begin{center}
\begin{tabular}[c]{|c|c|c|c|c|c|c|c|c|}
\hline 
Rx  & Horiz. Tx  & Meas. ToF, & Meas. ToF, & Ray-trace ToF & Ray-trace ToF & Ray-trace ToF & FDTD ToF & FDTD ToF \\
Depth  & Dist. & 1st Pulse &  2nd  Pulse & (Bulk Prop.)  & (Horiz. Prop.)  & (Surf. Prop.)  & 1st Pulse & 2nd Pulse \\
(ft.) & (ft.) & (ns) &  (ns) & (ns) &  (ns) &  (ns) & (ns) & (ns)\\

\hline
\multicolumn{9}{c}{Non-Shadowed Region in All Firn Models}\\
\hline
150 & 200 & 388 & -- & $377\pm1$ & -- & $391\pm7$ & $379\pm1$ & --\\
200 & 200 & 441 & -- & $438\pm2$ & -- & $446\pm2$  & $440\pm1$ &--\\
400 & 200 & 755 & -- & $737\pm3$ & -- & --  & $742\pm4$ &--\\
580 & 200 & 1072 & -- & $1038\pm3$ & -- & --  & $1053\pm4$ &--\\
400 & 480 & 1035 & -- & $1025\pm3$ & -- & $1054\pm13$  & $1033\pm5$ &--\\
580 & 480 & 1295 & -- & $1272\pm4$ & -- & $1287\pm3$  & $1285\pm5$ & --\\
\hline
\multicolumn{9}{c}{Edge of the Shadowed Region}\\
\hline
200 & 480 & 806 & -- & $782\pm7$  & $782\pm7$ & $809\pm27$  &  $808 \pm 7$ & -- \\
\hline
\multicolumn{9}{c}{Shadowed Region in All Firn Models}\\
\hline
200 & 1000 & 1315 & 1560 & -- & $1442\pm33$ & $1484\pm74$  & $1259\pm1$ & $1562\pm36$\\
400 & 1000 & 1601 & 1790 & -- & $1682\pm20$ & $1729\pm60$  & $1545\pm4$ & $1786\pm54$\\
600 & 1050 & 1873 & 2010 & -- & $1998\pm12$ & $2052\pm53$  & $1893\pm4$ & $2051\pm35$ \\
\hline
\end{tabular}
\end{center}
\caption[]{The measured and expected time of flight for a variety of experimental configurations,
  ray-tracing hypotheses, and FDTD simulations. For the ray-tracing hypotheses and FDTD simulations, we use two
  different firn models, which are based on the Hawley~\cite{hawley} and Alley~\cite{alley} data, 
  and are discussed in Section~\ref{sec:firnModel}.  We 
  report the mean of the results for each firn model and the standard deviation as an uncertainty.
  The time of flight shown for the data and the FDTD simulations 
  is the arrival of the maximum amplitude of the signal. 
  For the 600~ft. depth and 1050~ft. distance measurement, we report an uncertainty on the second pulse's arrival time
  that accounts for the observation of multiple peaks of comparable amplitude.
  In both the shadowed and the non-shadowed region,
  the data is more consistent with the results of the FDTD simulations than the ray tracer. 
  In the non-shadowed region, the data matches the predicted time of flight well, with $\sim1$\% agreement with the 
  FDTD simulations and $\sim3$\% agreement to the bulk propagation ray-tracing mode.
  We observe signals in the shadowed
  region, and the timing of the first pulse is consistent with FDTD simulations to within 3\% and differs from
  the ray tracer prediction for horizontal propagation by 9\%.
  \label{tab:times} 
 }
\end{table*} 
\begin{table*}
\begin{center}
\begin{tabular}[c]{|c|c|c|c|c|c|}
\hline 
Rx  & Horiz. Tx  & Measured   & Relative & Relative E-field & Tx Angle\\
Depth (ft.) & Distance (ft.) & Voltage  (mV) & Voltage (Arbitrary) &FDTD Simulation (Arbitrary) & (Degrees)\\
\hline
\multicolumn{6}{c}{Non-Shadowed Region in All Firn Models}\\
\hline
150 & 200 & 776  & 1.00 & $1.00\pm0.30$ & 17\\ %
200 & 200 & 627  & 0.81 &$0.71\pm0.09$ & 28\\ %
400 & 200 & 431  & 0.56 &$0.12\pm0.01$ & 54\\ %
580 & 200 & 137 & 0.18 &$0.04\pm0.001$ & 64\\ %
400 & 480 & 182  & 0.23 &$0.63\pm0.23$ & 8\\%
580 & 480 & 236  & 0.30 &$0.13\pm0.15$ & 30\\ %
\hline
\multicolumn{6}{c}{Edge of the Shadowed Region}\\
\hline
200 & 480 & 59 & 0.076 &$0.07\pm0.01$ & 0 \\ %
\hline
\multicolumn{6}{c}{Shadowed Region in All Firn Models}\\
\hline
200 & 1000 & 1.3  & $1.7\times10^{-3}$ &$4.8\times10^{-4}\pm1\times10^{-5}$  & 0 \\ %
400 & 1000 & 1.5  & $1.9\times10^{-3}$ &$5.0\times10^{-4}\pm1\times10^{-5}$ & 0 \\ %
600 & 1050 & 2.1  & $2.7\times10^{-3}$ &$5.0\times10^{-4}\pm1\times10^{-5}$ & 0 \\ %
\hline
\end{tabular}
\end{center}
\caption[]{The measured signal peak-to-peak amplitudes of the first received signal for a variety of 
  experimental configurations (receiver depth and transmitter horizontal distance), 
  after accounting for relative attenuation in the system between different
  configurations.  Signals
  in the shadowed region are uniformly smaller than signals in the non-shadowed
  region, even after accounting for signal path length differences. 
  The fourth column shows the measured voltages, normalized to the first row, for direct comparison
  to the values in the fifth column.
  The fifth column shows the amplitude of the first signal
  predicted by the FDTD simulations assuming firn models based on the Hawley data~\cite{hawley} and the Alley 
  ice core data~\cite{alley} (discussed in Section~\ref{sec:firnModel}).  We report the mean of the simulation results
  using the two firn models and the standard deviation as an uncertainty.
  The simulation assumes a dipole transmitter and an isotropic receiver and does not include
  any electric field loss from attenuation in the ice.  The amplitudes
  predicted from the FDTD simulation show significant suppression at steep angles 
  due to the beam pattern of the simulated transmitter (most prominent at the 580~ft. depth, 200~ft. horizontal distance 
  geometry), which would not appear in the data, since we optimized the transmitter angle
  for the measurements.  The predicted elevation angle (with respect to horizontal) of the transmission from the
  ray tracer is shown in the last column.
  \label{tab:amplitudes} 
 }
\end{table*} 
\begin{figure}[]
      \begin{center}
        \includegraphics[width=8cm]{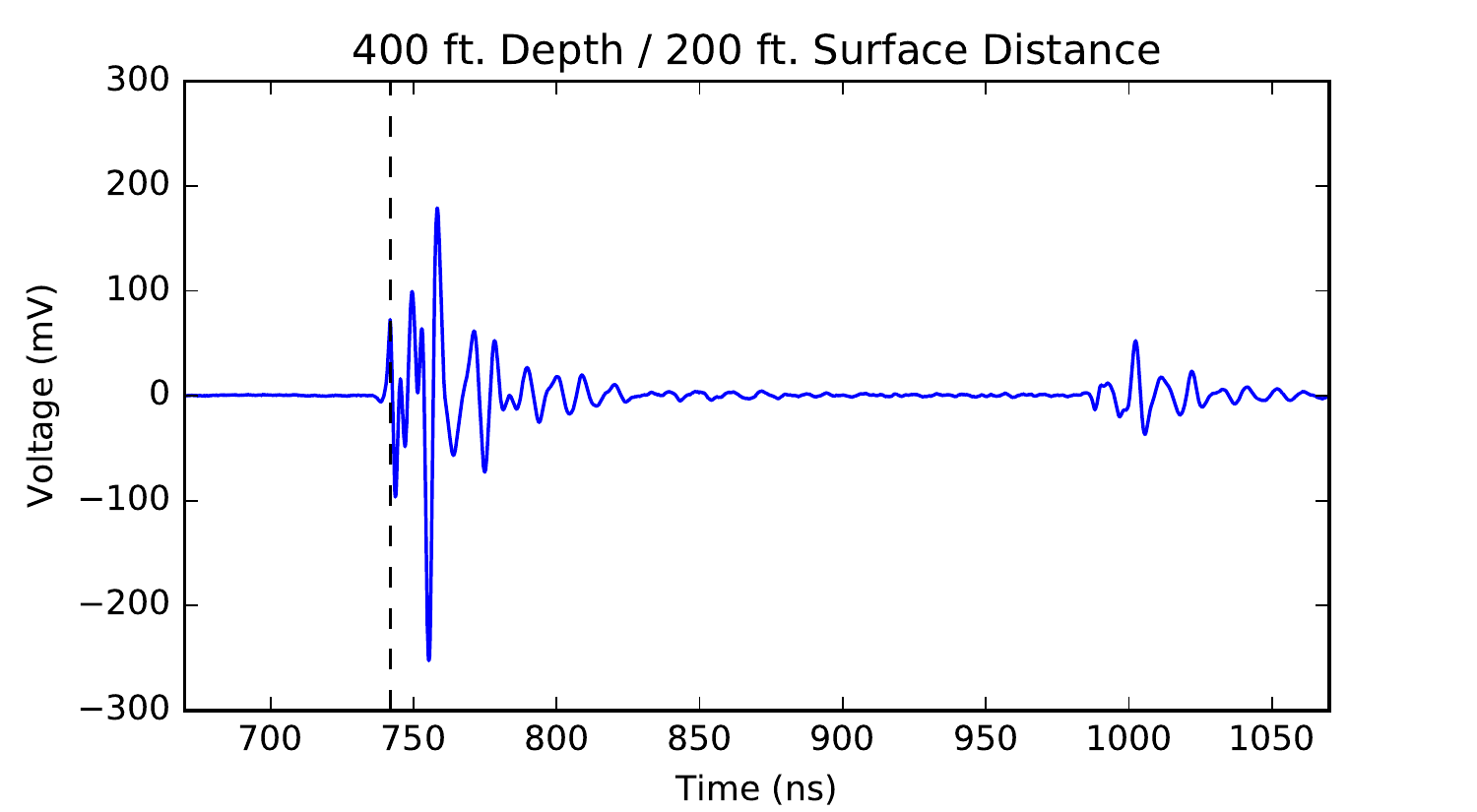}
        \includegraphics[width=8cm]{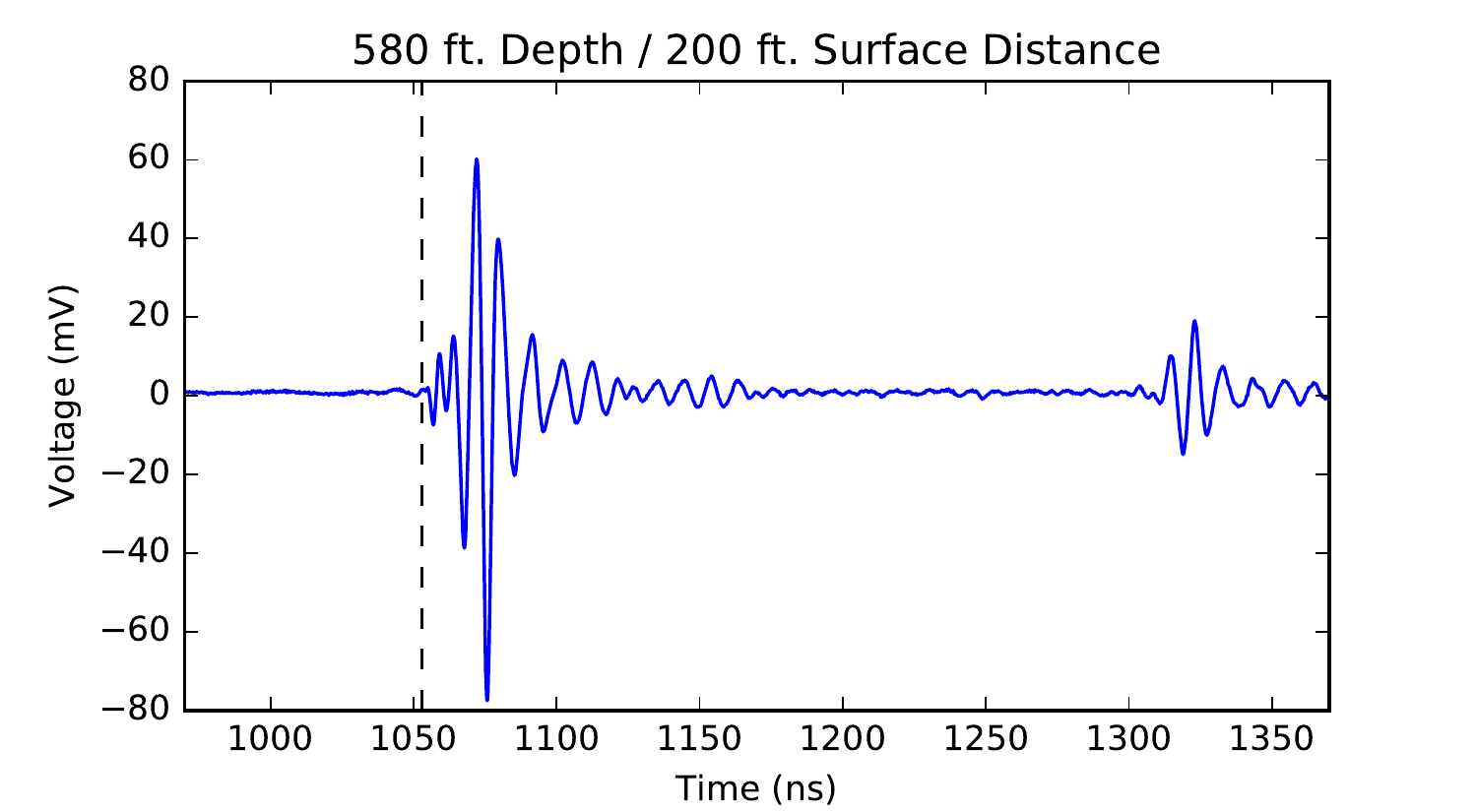}\\
        \includegraphics[width=8cm]{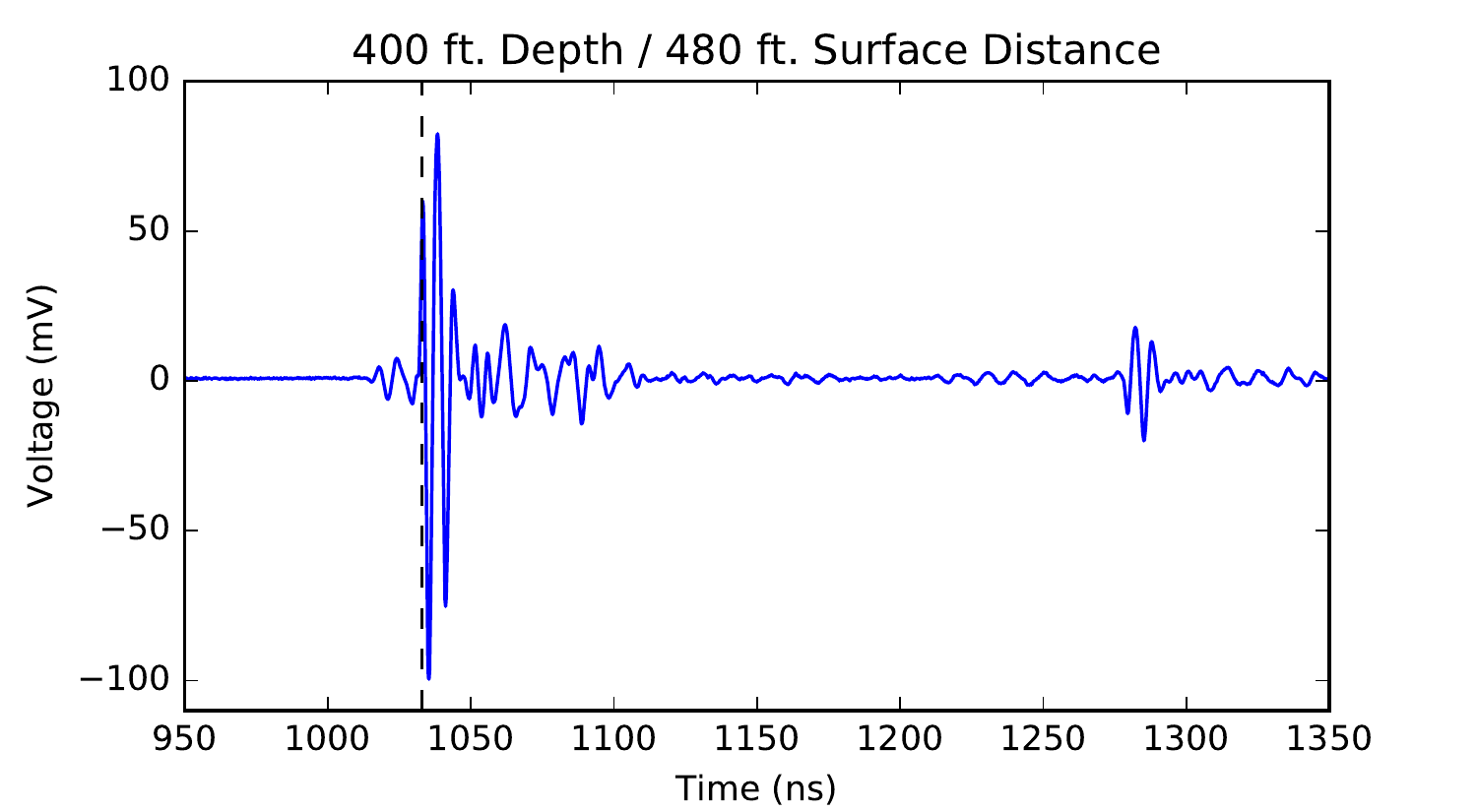}
        \includegraphics[width=8cm]{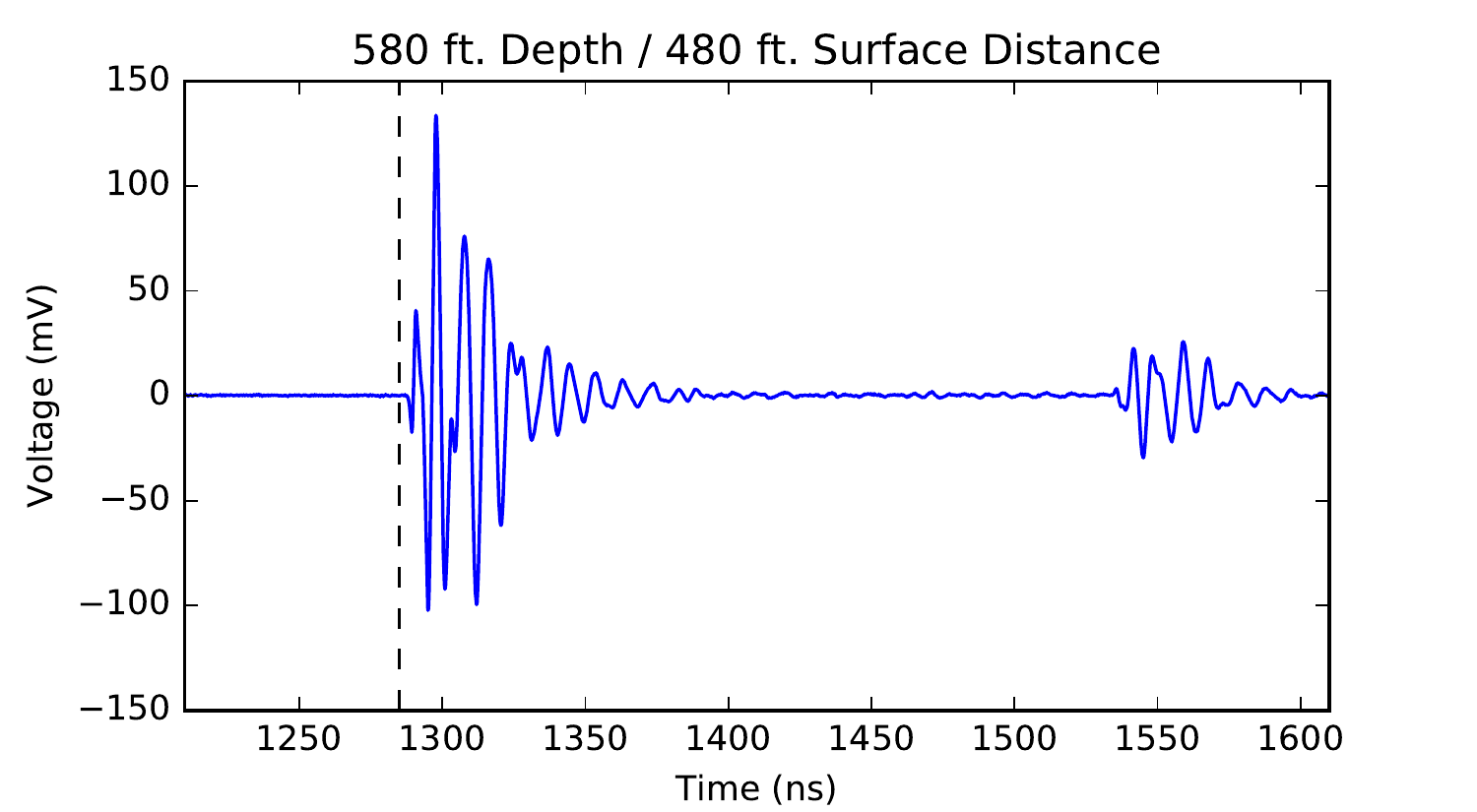}
    \end{center}
    \caption{Top two: Example waveforms from locations 
      where there is only a bulk propagation ray-bending solution.  Bottom two: Example waveforms from locations
      where there is a bulk propagation ray-bending solution and an additional surface-wave solution.  The time shown
      is the absolute propagation time of the signal after removing the system delay.  The dashed line
      shows the predicted time of flight from the FDTD simulations, shown in Table~\ref{tab:times}.  
      The second pulse seen at 245~ns
      after the initial pulse is a known reflection in the system corresponding the out-and-back time along one section 
      of LMR-600 cable, and appears in all data.  There is no evidence
      that a large fraction of power is contained in
      an additional horizontally-propagating surface wave for the waveforms shown in the bottom panels, 
      which would arrive up to 20~ns after the first signal 
      (times shown in Table~\ref{tab:times}). } 
    \label{fig:allowedWaveform}
  \end{figure}

Table~\ref{tab:times} shows the measured time of flight of the signals observed compared to the expected time of
flight from the three modes described in Section~\ref{sec:models} and 
the simulated time of flight from the FDTD simulations described in Section~\ref{sec:sims}.  
We estimate that we measure the transmitter
distance across the surface to within 3~ft. and the receiver depth down the borehole to within 1~ft., using a measuring
tape. We include the time of flight for the first observed pulse, 
and for cases where we saw a second pulse (separate from the known reflection in the system), we report
its time of flight as well. Note that for each geometry, only one or two modes out of the three 
in the ray tracer converge on a solution.
In the non-shadowed region, the time of flight matches the FDTD simulation results within 1\%.  In the shadowed
region, the time of flight is uniformly less consistent with the predictions. The uncertainty on the predicted
time of flights from 
using different firn models is larger for the second pulse.  This is consistent with the first pulse's 
surface wave origin, and the second pulse originating from horizontal propagation along layers in the firn, which
change from model to model.  The measured arrival times match the FDTD simulation results in the 
shadowed region to within 3\%.

\begin{figure}[]
      \begin{center}
        \includegraphics[width=8cm]{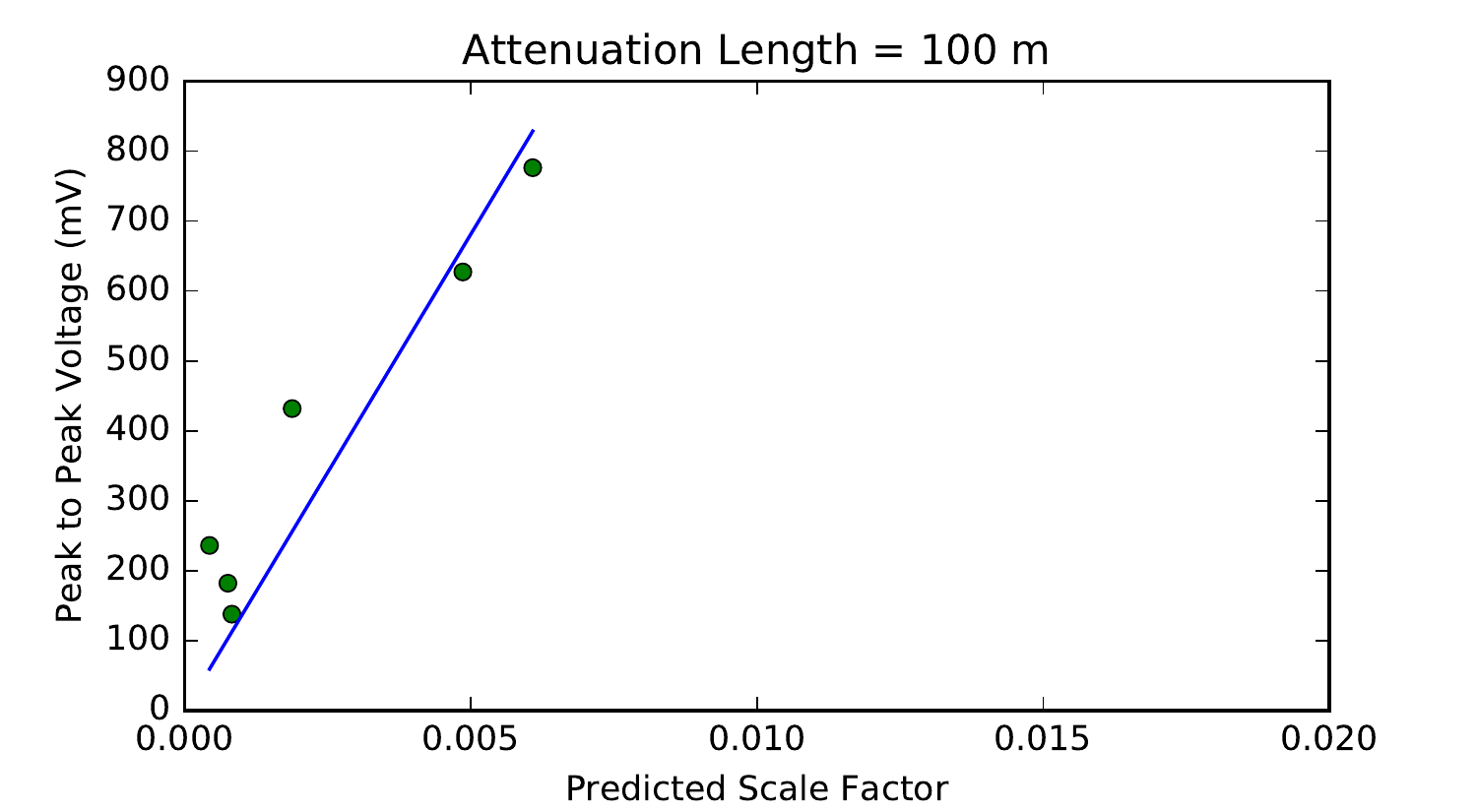}
        \includegraphics[width=8cm]{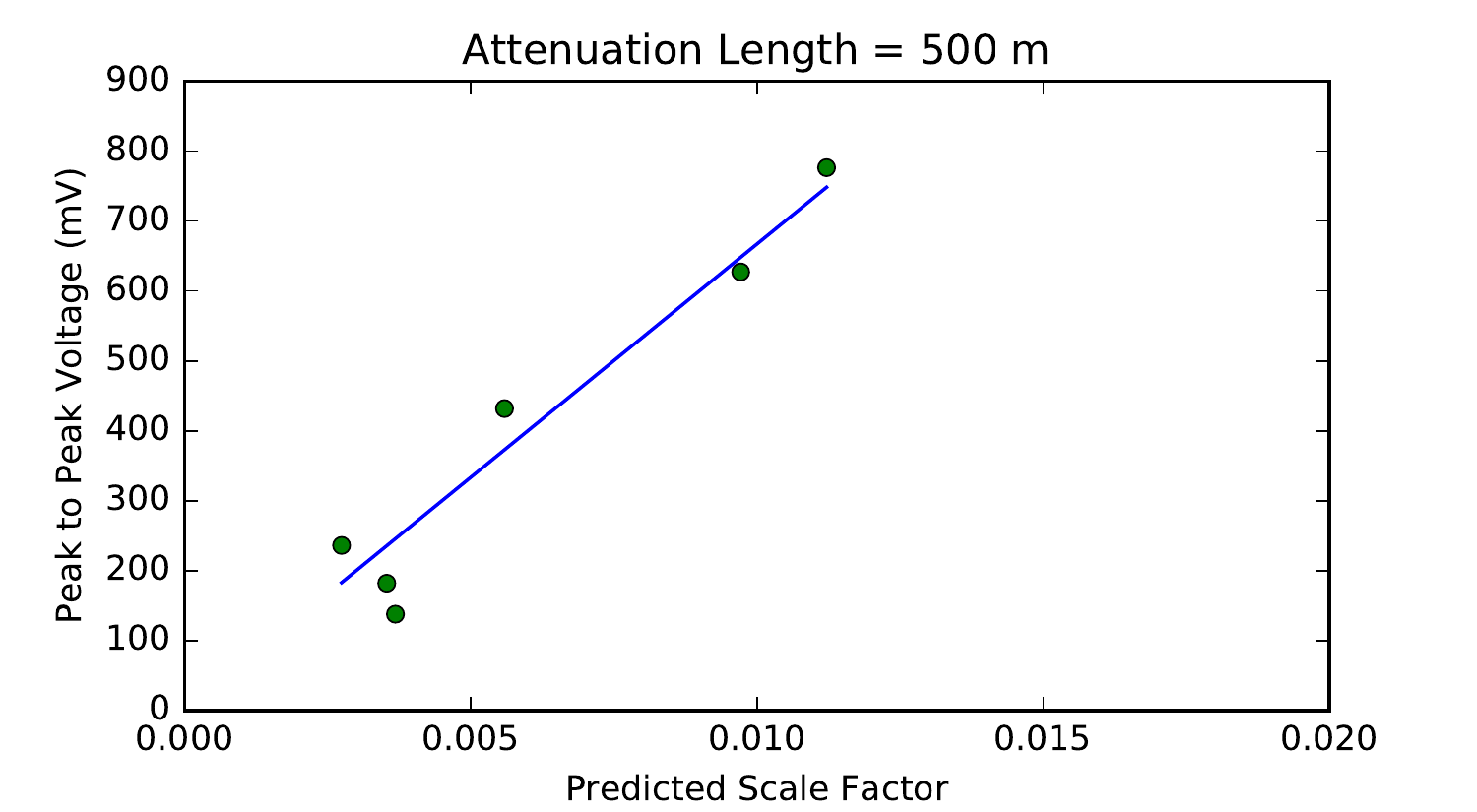}\\
        \includegraphics[width=8cm]{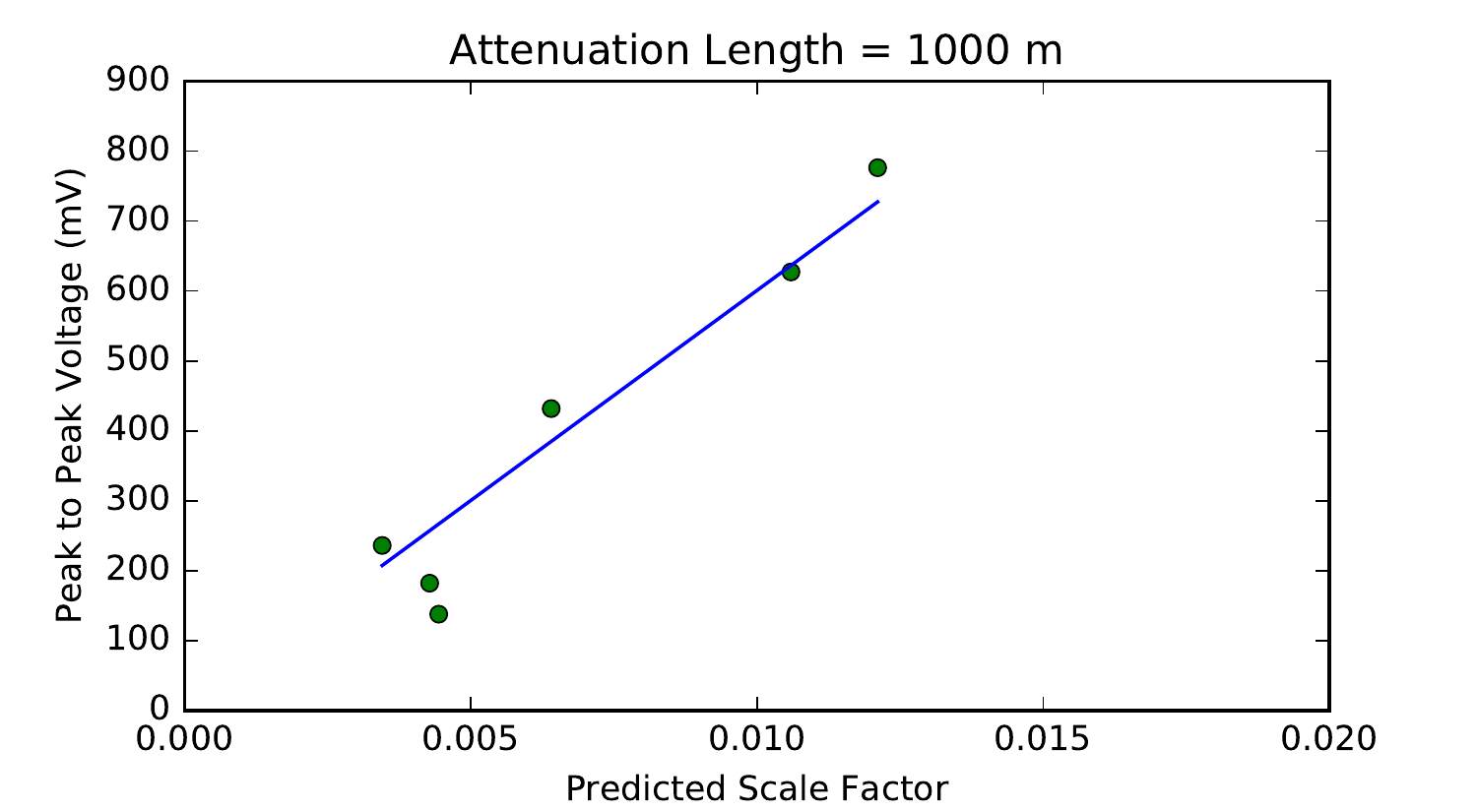}
        \includegraphics[width=8cm]{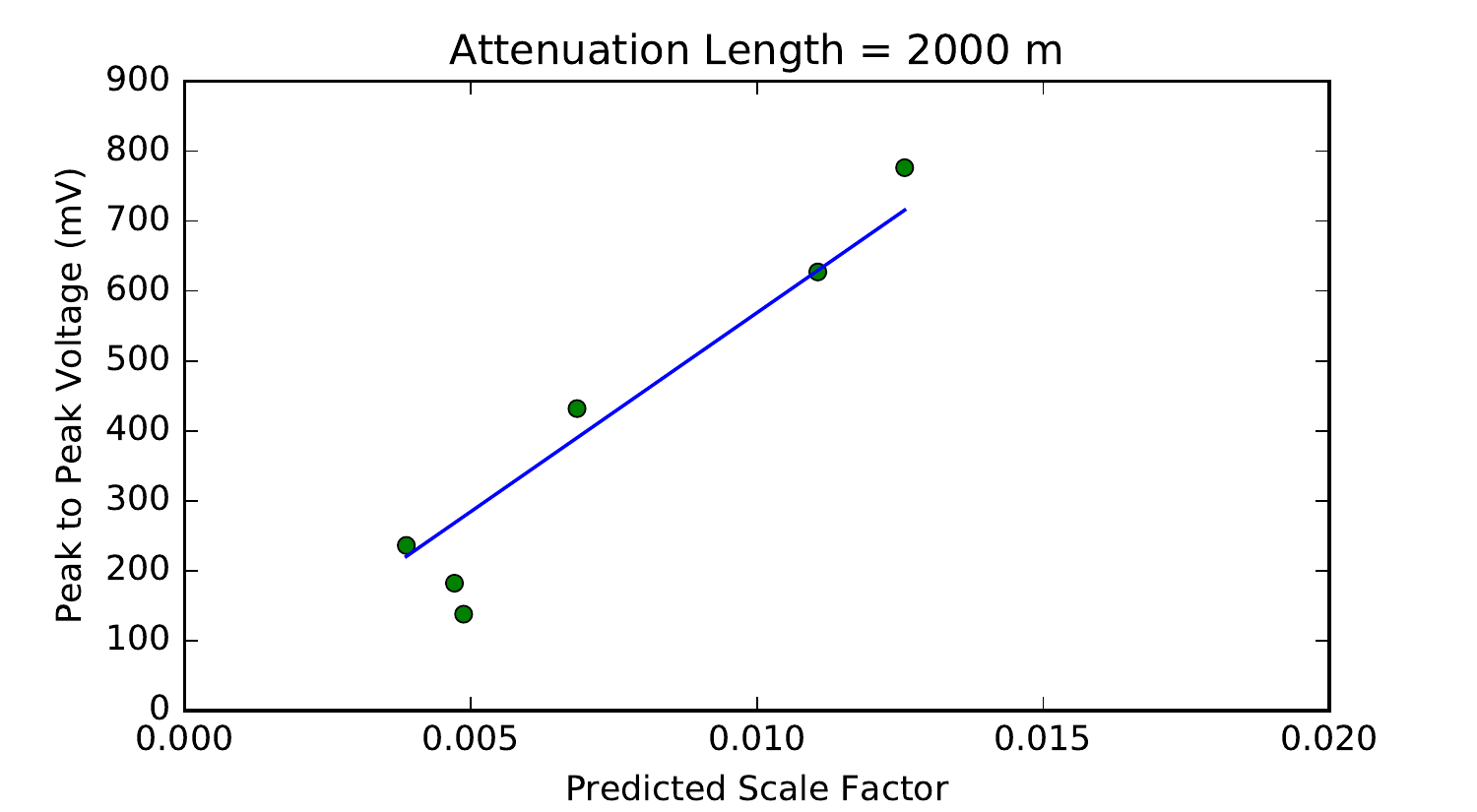}
    \end{center}
    \caption{Green points: The peak-to-peak voltage measured for the largest (and first to arrive in all cases) 
      signal in the 
      non-shadowed region vs. a predicted scale factor compared to a receiver at 1~m 
      for a given field attenuation length and 
      assuming $1/r$ scaling for electric field.  
      Blue line: A least squares linear fit to the data, given the assumed attenuation length and $1/r$ 
      scaling in each case. 
      The slope of this line is used to derive the presumed electric field at 1~m from the 
      transmitter in each case and errors corresponding to the full spread of values are derived from
      fitting a line to only one measurement at a time.  The results are shown in Table~\ref{tab:efield1m}.
    } 
    \label{fig:allowedOneOverR}
  \end{figure}
Table~\ref{tab:amplitudes} shows the measured amplitude (peak-to-peak voltage) of the signals seen for each geometry
and the predicted amplitudes from the FDTD simulation.  The FDTD simulation assumes a dipole transmitter 
and an isotropic receiver and does not include any electric field loss from attenuation in the ice.  At the 
largest distance reported, the path length is smaller than one $e$-folding, assuming previously-reported
attenuation lengths at Summit Station~\cite{avva, macgregor}.
The signals at the 1000~ft. and 1050~ft. horizontal transmission distance were so weak that we removed a 20~dB attenuator
from the system in order to see the signals.  This has already been accounted for in the numbers reported in this table.
Note that the amplitudes measured in the shadowed region are all smaller than those measured in the 
non-shadowed region, even after accounting for relative path lengths through the ice.  The amplitudes
predicted from the FDTD simulation show significant suppression at steep angles 
due to the beam pattern of the simulated transmitter (most prominent at the 580~ft. depth, 200~ft. horizontal distance 
geometry), which would not appear in the data, since we optimized the transmitter angle
for the measurements.

\subsection{Measurements in the Non-Shadowed Region}

We took data in six configurations in the non-shadowed region, and the time of flights and amplitudes are 
shown in Tables~\ref{tab:times} and~\ref{tab:amplitudes}.  Figure~\ref{fig:allowedWaveform} shows example 
waveforms from four of the geometries, demonstrating the range of 
received signal shapes in the non-shadowed region.  
In all cases in the non-shadowed region, the waveforms show a clear initial impulse, followed
by the smaller pulse from the reflection in the system 245~ns later.  
In two of the six geometries, only one ray-tracing mode (the bulk propagation mode) converges to a solution
using the ray tracer, whereas in the other 
four, two modes (the bulk propagation and surface-wave modes) converge, predicting that a surface ray should arrive up to 
20~ns after the direct ray 
(see Table~\ref{tab:times}).  The top two panels in Figure~\ref{fig:allowedWaveform}
show two cases where there is only a bulk propagation solution, and the bottom two panels show two cases where
there is an additional surface-wave solution.  There is no evidence that a large fraction of power 
is contained in such a horizontally-propagating surface wave, since there is no clear distinction
between the top and bottom panels in terms of signal shape.  The frequency of the signals is lower
than that observed in air, as expected, with -3~dB points at $\sim$90--150~MHz, but with
significant power out to 220 MHz. 

We test consistency between the bulk 
propagation (ray-bending) model and the measured 
amplitudes in this region.  Figure~\ref{fig:allowedOneOverR} shows the amplitude data plotted against
the predicted scale factor compared to a receiver at 1~m 
using $1/r$ scaling for electric field, and a variety of choices of bulk
field attenuation length.  The scale factor is given by:

\begin{equation}
\frac{1~\mathrm{m}}{r}\times e^{-r/l_{\mathrm{bulk}}},
\end{equation}
where $r$ is the path length in meters and $l_{\mathrm{bulk}}$ is the field attenuation length for bulk propagation.

Ideally, for the best choice of attenuation length and field propagation behavior, 
the data would fall on a line with a y-intercept of zero.
The slope of the line indicates the overall strength of the transmitter of the system, giving the 
equivalent voltage received if the transmitter and receiver were placed 1~m apart.  
The slopes for each choice of attenuation length are shown in Table~\ref{tab:efield1m}.  
The data do not fall directly on a straight line, presumably due to variability
of the snow, the achieved coupling at each location, and interference effects 
(see Figures~\ref{fig:fdtdHawley} and ~\ref{fig:fdtdAlley}), so we assign uncertainties in Table~\ref{tab:efield1m} 
that correspond to the full spread of values calculated using only one of the six measurements at a time.  %
\begin{table}[h]
\begin{center}
\begin{tabular}[c]{|c|c|}
\hline 
Bulk Field  & Derived Voltage  \\
Attenuation Length (m) & Received at 1~m (V) \\
\hline
& \\[-1.5ex]
100 & $136^{+406}_{-9}$\\[1ex]
500 & $67^{+19}_{-29}$ \\[1ex]
\bf{1000} & $\mathbf{60^{+8}_{-29}}$ \\[1ex]
2000 & $57^{+6}_{-29}$ \\[1ex]
\hline
\end{tabular}
\end{center}
\caption[]{The derived voltage received for a receiver 1~m away 
  from our transmitter, given a variety
  of choices of bulk field attenuation length at Summit Station.  Previous measurements show that 
  the bulk field attenuation 
  length at 300~MHz is $\sim1000$~m, and we have
  highlighted the row in the table that corresponds to that value.  The uncertainties correspond to the 
  full spread of values calculated using only one of the six measurements at a time.
  \label{tab:efield1m} 
 }
\end{table} 

Although this data set is poor for determining the attenuation length 
due to the relatively short baselines probed, previous measurements show that the bulk field attenuation 
length at 300~MHz is $1022^{230}_{-253}$~m over the upper 1500~m of ice at Summit Station~\cite{avva}, also consistent 
with~\cite{macgregor}.  We therefore highlight the row in Table~\ref{tab:efield1m} that 
most closely corresponds to this value.
\begin{figure}[]
      \begin{center}
        \includegraphics[width=8.5cm]{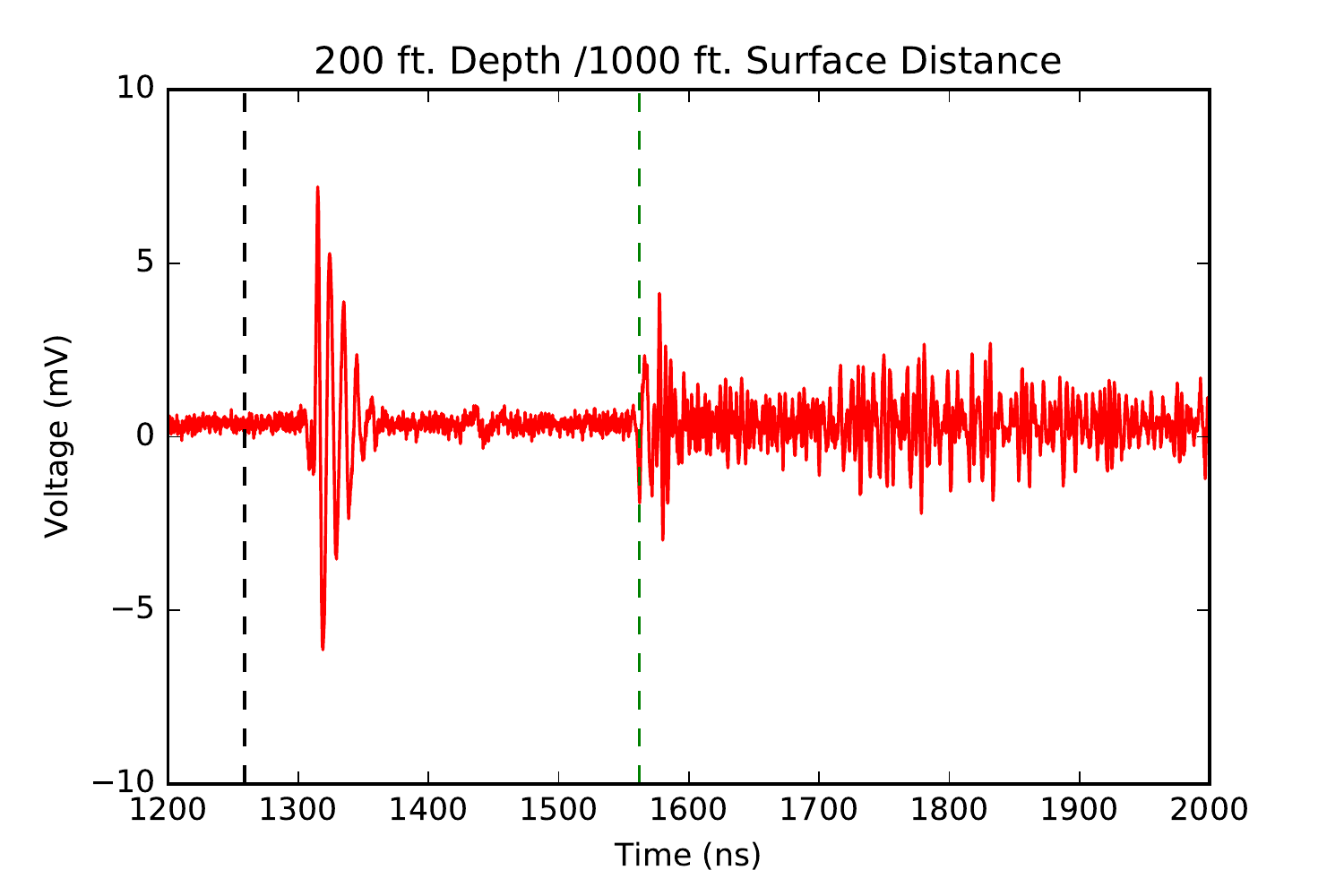}
        \includegraphics[width=8.5cm]{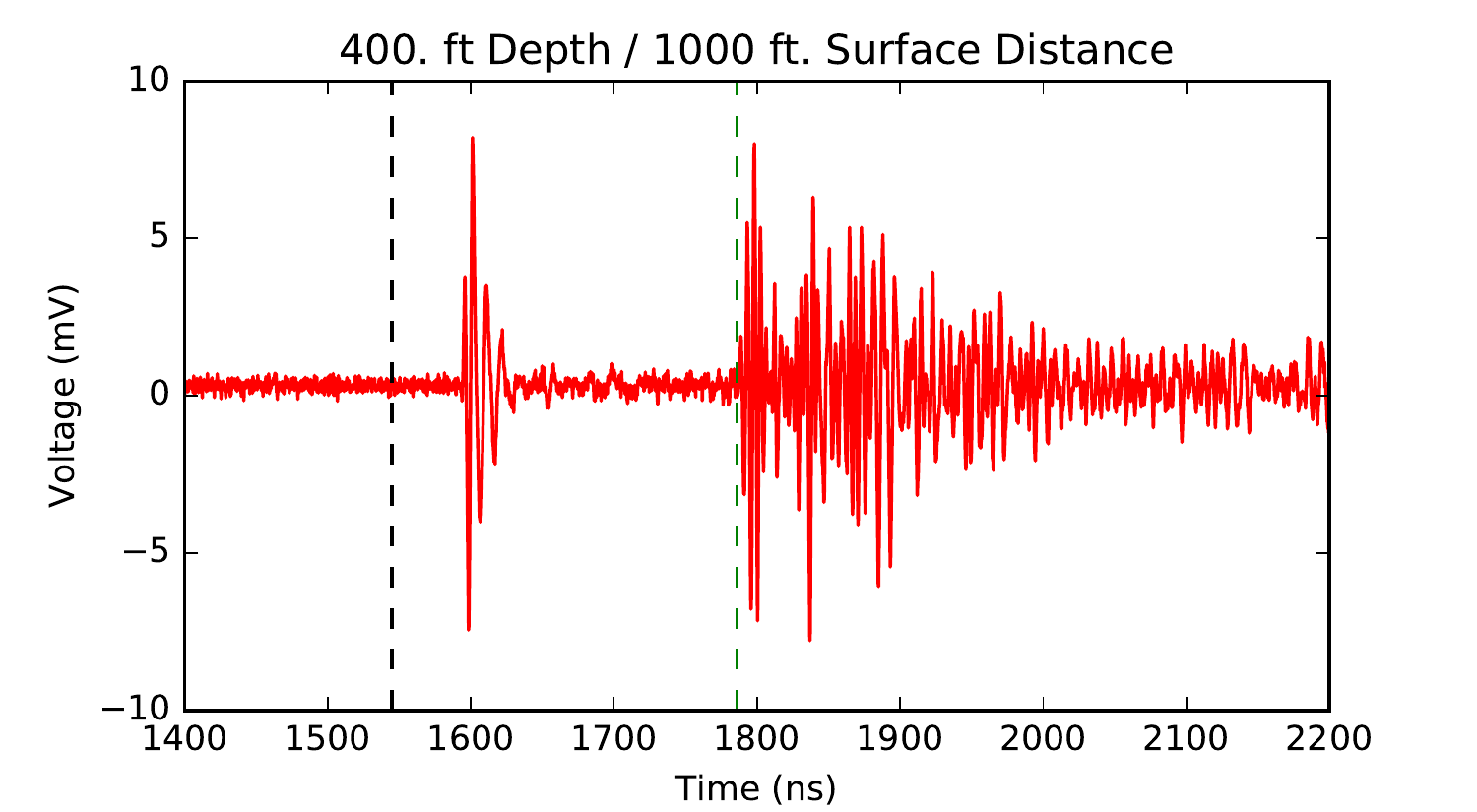}
        \includegraphics[width=8.5cm]{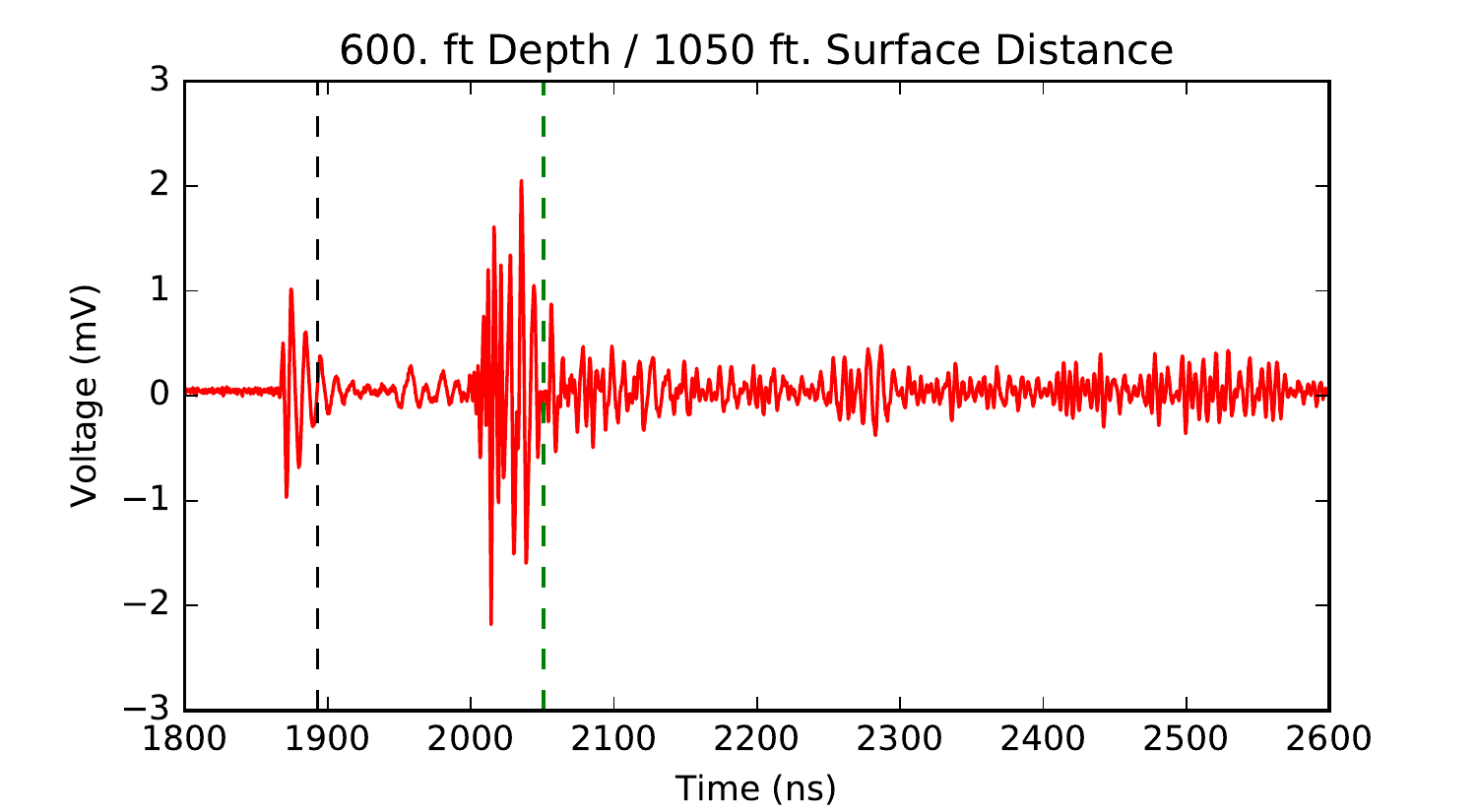}
    \end{center}
    \caption{All three waveforms received in the shadowed region after accounting for 
      relative attenuation in the system among measurements.  Note that there are two distinct signals,
      of varying relative amplitudes between the first and second main signal, and there is also significant power seen
      after the main signals, which was never observed in the non-shadowed region. The black dashed line
      shows the predicted time of flight of the first pulse 
      and the green dashed line  
      shows the mean predicted time of flight
      of the second pulse from the FDTD simulations, shown in Table~\ref{tab:times}.  
      Note that the timescale is double that in Figures~\ref{fig:airWaveform} and \ref{fig:allowedWaveform}.  
      The characteristics that distinguish these signals from those seen in the non-shadowed region (the changes
      in relative amplitude between the first and second pulse as a function of depth, the signal shape differences between
      the first and second pulses, and the additional power that follows the second pulse) are
      directly seen in the FDTD simulations of the shadowed region, 
      shown in Figures~\ref{fig:fdtdHawley} and~\ref{fig:fdtdAlley}.
    } 
    \label{fig:shadowedWaveform}
  \end{figure}
\begin{figure}[]
      \begin{center}
        \includegraphics[width=8.5cm]{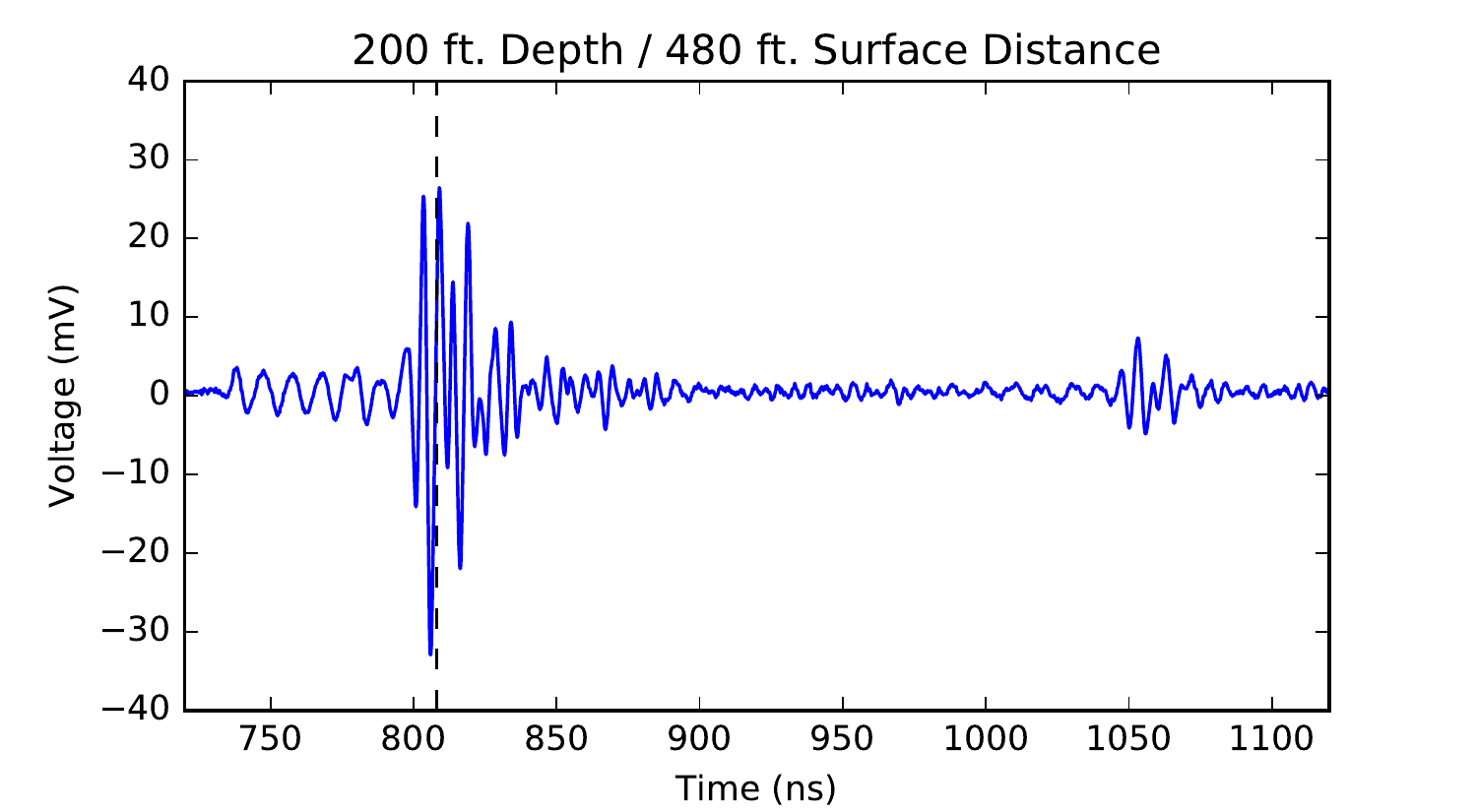}
    \end{center}
    \caption{The waveform received for the geometry predicted to be near the edge of the shadowed region.  In some
      firn models it is in the shadowed region and in other models it is in the non-shadowed region.   The dashed line
      shows the predicted time of flight from the FDTD simulations, shown in Table~\ref{tab:times}. The 
      waveform shape and
      characteristics are consistent with measurements from the non-shadowed region (see 
      Figure~\ref{fig:allowedWaveform}), but the 
      amplitude is somewhat suppressed compared to the other measurements in the non-shadowed region.  
    } 
    \label{fig:shadowedWaveformMaybe}
  \end{figure}

\subsection{Measurements in the Shadowed Region}
We also made measurements in three configurations that are shadowed from bulk ray-bending propagation.  
Figure~\ref{fig:shadowedWaveform} shows the data from these configurations.   
Counter to a bulk-propagation-only model, 
we observe signals in all three configurations
in the shadowed region, consistent with previous measurements~\cite{barwick} and also consistent with
FDTD simulations.  

The amplitudes of these signals are uniformly smaller
than the signals observed in the non-shadowed region, as shown in Table~\ref{tab:amplitudes}.
The waveforms all show a different set of characteristics from those in the non-shadowed region,
and are remarkably consistent with the predicted waveforms from the FDTD simulations 
(see Figures~\ref{fig:fdtdHawley} and~\ref{fig:fdtdAlley}), both in signal shape and in amplitude 
relative to the non-shadowed region.
There are two main pulses seen in all cases in this region, but never seen in the non-shadowed region.  
The relative amplitude of these two pulses changes as a function of depth, consistent 
with FDTD simulations.  Moreover, the first pulse has a clean signal shape, whereas the second pulse 
is not as clean, and there is significant power that follows the second pulse for hundreds of nanoseconds.  This is also
consistent with the behavior observed in the simulations.

In addition to the three measurement configurations that are comfortably in the shadow, there is one measurement 
configuration that according to 
the FDTD simulations and the ray tracer resides at the edge of the shadowed region (depending on the firn model parameters
one chooses, this configuration either lies inside or outside the shadowed region).  The waveform for this configuration
is shown in Figure~\ref{fig:shadowedWaveformMaybe}.  The waveform is consistent in its characteristics with other waveforms 
in the non-shadowed region, but the amplitude is somewhat suppressed, consistent with FDTD simulations using the 
model based on the Hawley~\cite{hawley} data and the Alley data~\cite{alley}.

Although there is no ray-bending bulk propagation ray-tracing solution for these geometries, the 
surface-wave and horizontal
propagation modes~\cite{ralston, barwick} in the ray tracer 
and the FDTD simulation predict signals in this region.  The 
measured time differences between the first and second pulses
are shown in Table~\ref{tab:times}.  These times are consistent with the predictions 
from FDTD simulations (within 3\%), 
and less so with the ray tracer predictions for horizontal and surface-wave modes.

\subsection{Coupling into a Horizontally-Propagating Mode}
\label{sec:coupling}
\begin{table*}
\begin{center}
\begin{tabular}[c]{|c|c|c|c|}
\hline 
Bulk Field & Horizontal Tunnel Field & Propagation Loss & Derived Electric Field  \\
Attenuation Length (m) & Attenuation Length (m) &  & Coupling Fraction  \\
\hline 
& & & \\[-1.5ex]
100 & 100 & $1/\sqrt{r}$ & $0.09^{+0.42}_{-0.04}$ \\[1ex]
100 & 100 & $1/r$ & $0.15^{+0.46}_{-0.07}$  \\[1ex]
500 & 500 & $1/\sqrt{r}$ & $0.014^{+0.031}_{-0.007}$\\[1ex]
500 & 500 & $1/r$ & $0.024^{+0.031}_{-0.011}$  \\[1ex]
1000 & 1000 & $1/\sqrt{r}$ & $0.011^{+0.023}_{-0.005}$\\[1ex]
1000 & 1000 & $1/r$ & $0.019^{+0.017}_{-0.009}$  \\[1ex]
2000 & 2000 & $1/\sqrt{r}$ & $0.010^{+0.020}_{-0.005}$ \\[1ex]
2000 & 2000 & $1/r$ & $0.017^{+0.019}_{-0.008}$  \\[1ex]
1000 & 500 & $1/\sqrt{r}$ & $0.014^{+0.024}_{-0.007}$\\[1ex]
\bf{1000} & \bf{500} & \bf{$\mathbf{1/r}$} & \bf{$\mathbf{0.023^{+0.023}_{-0.010}}$}\\[1ex]
\hline
\end{tabular}
\end{center}
\caption[]{The calculated electric field coupling fraction under a variety of assumptions, including a choice of 
  $1/r$ or $1/\sqrt{r}$ propagation loss, and various values of the bulk and horizontal tunnel field attenuation lengths.
  A previous study has determined the bulk field attenuation length 
  at Summit Station to be $\sim1000$~m~\cite{avva}, and separate studies at South Pole and Moore's Bay in Antarctica
  have shown results consistent with a 500~m field attenuation length and $1/r$ propagation for a horizontal 
  mode~\cite{barwick}.  The row consistent with these previous measurements is highlighted.  
  \label{tab:coupling} 
 }
\end{table*} 

For a wave to transition from moving through the bulk medium to propagating along
a horizontal or surface-wave mode, power must couple from one mode to the other, rather than continuing 
the bulk propagation where it would either refract through the surface or reflect off the surface.
We can use our data to derive the coupling fraction of power into a horizontally-propagating 
or surface-wave mode for signals propagating from deep in the ice.

We begin with the calculation of the signal strength for a hypothetical receiver placed 1~m from the transmitter,
shown in Table~\ref{tab:efield1m}.  This calculated ``system amplitude'' depends on the choice of bulk attenuation 
length.  Using this system amplitude for a range of bulk attenuation length choices, 
and testing a variety of choices of attenuation length for the horizontally-propagating mode 
(which can in principle be different from the bulk attenuation length)
and whether the electric field
in the horizontal mode falls as $1/r$ or $1/\sqrt{r}$, 
we calculate the coupling fraction in each case using the following relationships
for the received voltage ($V_\mathrm{r}$):

\begin{equation}
V_{\mathrm{r}} = \frac{V_{1\mathrm{m}} \times f \times 1~\mathrm{m}}{r_{\mathrm{bulk}}+r_{\mathrm{horiz}}} e^{-r_{\mathrm{bulk}}/l_{\mathrm{bulk}}} \times e^{-r_{\mathrm{horiz}}/l_{\mathrm{horiz}}} 
\label{eqn:oneoverr}
\end{equation}
or
\begin{equation}
V_{\mathrm{r}} =  \frac{V_{1\mathrm{m}} \times f \times 1~\mathrm{m}}{\sqrt{r_{\mathrm{bulk}}}\sqrt{r_{\mathrm{bulk}}+r_{\mathrm{horiz}}}} e^{-r_{\mathrm{bulk}}/l_{\mathrm{bulk}}} \times e^{-r_{\mathrm{horiz}}/l_{\mathrm{horiz}}},
\label{eqn:oneoversqrtr}
\end{equation}
where $V_{1\mathrm{m}}$ is the voltage that would be measured for a receiver at 1~m (shown in Table~\ref{tab:efield1m}),
$f$ is the electric field coupling fraction, $r_\mathrm{bulk}$ is the path length of bulk propagation in meters,
$r_\mathrm{horiz}$ is the path length of horizontal or surface-wave propagation in meters, $l_{\mathrm{bulk}}$ is the 
field attenuation length for bulk propagation, and $l_{\mathrm{horiz}}$ is the field attenuation length
for horizontal or surface-wave propagation.  The component of the loss from propagation effects (rather than 
attenuation) stated in 
Equations~\ref{eqn:oneoverr} and~\ref{eqn:oneoversqrtr} 
is not equivalent to taking the scaling factor $1/r$ or $1/\sqrt{r}$ for each segment (bulk and horizontal
propagation) and multiplying them together.  Simply multiplying the two factors would propagate a 
signal from a 1~m distance, and then propagate that same signal again from the same 1~m distance, which 
does not give the correct total scaling factor.  Rather,
we want to propagate the signal from a 1~m distance to the end of the first leg of the path, 
and then propagate the remaining signal through the second leg of the path, which results 
in Equations~\ref{eqn:oneoverr} and~\ref{eqn:oneoversqrtr}.

The results for a variety of parameter choices are shown in 
Table~\ref{tab:coupling}.  Uncertainties in the Table correspond to the full spread of values 
obtained if we fit only one of the three measurements at a time.  
We note that applying a beam pattern correction to the amplitude throughout the
analysis reduces the calculated coupling fraction in electric field, but by 
less than 20\% in all cases shown.

Based on previous measurements of bulk attenuation length 
at Summit Station and of horizontally-propagating modes at South Pole and Moore's Bay, 
we choose to highlight 
the row in Table~\ref{tab:coupling} that shows choices of parameters most consistent with those measurements
(a bulk field attenuation length at Summit Station of 1000~m, a horizontally-propagating field attenuation length
of 500~m, and a propagation loss of $1/r$ for the horizontal mode).
These parameters give a 2.3\% coupling fraction in electric field (0.05\% in power), 
with an uncertainty permitting values ranging from 1.3\% to 4.6\%, representing the full range
of values obtained by using only one of the three measurements at a time.  
For almost all choices of attenuation length,
the best-fit coupling fraction in electric field amplitude is 2.4\% (0.06\% in power) or less 
to explain the small amplitudes seen in the shadowed region.
The only exception is for extremely pessimistic attenuation lengths of 100~m both
in the bulk and for the horizontally-propagating mode, which predicts small signal strengths 
for all signals and has a worse fit to our data in the non-shadowed region.  In this case,
and assuming $1/r$ propagation, the coupling fraction in electric field is 15\% 
(corresponding to 2.3\% in power). %

\section{Discussion}
\label{sec:discussion}

Although we observe signals in the shadowed region, the amplitudes of these signals 
are uniformly smaller than those observed in the non-shadowed region,
even after accounting for signal path length differences and attenuation.  Our data is remarkably
consistent with FDTD simulations of the as-measured firn at Summit Station.  
The measured time of flights are consistent, both in the shadowed and non-shadowed region, with
the simulations.  The waveform shapes are also consistent, which is especially interesting in the 
shadowed region.  The relative 
amplitudes predicted by FDTD simulations are also consistent in their order of magnitude. 
Given the small observed amplitudes
in the shadowed region, we find a best-fit electric field coupling fraction of 
2.4\% or less (0.06\% in power), representing the 
electric field that is coupled into a horizontally-propagating mode, rather than reflected, refracted, 
or bent. The amplitudes of signals in the shadowed region are shown to be significantly smaller than those in the
non-shadowed region, both in the data and in all FDTD simulations.

Knowledge of the relative amplitudes of signals in
the shadowed region compared to the non-shadowed region
is important for estimating the effective volume of experiments that aim to detect radio 
emission from neutrinos interacting in glacial ice, since the radio emission from neutrinos
would travel at a variety of angles through the firn.  
Our FDTD simulations show that the exact choice of firn model has significant impact on the 
exact amplitudes, signal shapes, and signal paths for a certain geometry, especially in the shadowed
region.  This is not surprising, since propagation in that region (other than surface propagation)
arises from density perturbations in the firn, which vary from model to model.  This presents a challenge
for event reconstruction and volumetric acceptance calculations, especially 
for near-surface detectors, since the firn near the surface changes most dramatically from year to year and 
on short distance scales.  The relative amplitudes of signals in the shadowed compared to the non-shadowed
region could potentially be different at different sites, such as South Pole, which has yet 
a different set of exact density perturbations in the firn.

The observation of multiple signals (e.g. a surface wave and a bulk-propagation mode in the non-shadowed
region or a surface wave and a horizontal mode in the shadowed region) from a 
single neutrino event would help determine the vertex of the event, relevant for
energy and angular reconstruction.

\section{Acknowledgements}
We would like to thank CH2M Hill and the US National 
Science Foundation (NSF) for the dedicated, knowledgeable, and
extremely helpful logistical support team enabling us to perform our work at Summit Station, 
particularly to J. Jenkins. We are deeply indebted to those who dedicate their
careers to help make our science possible in such remote environments.
This work was supported by the Kavli Institute for Cosmological Physics at the University of
Chicago, Department of Energy Award DE-SC0009937, NSF Award~1752922 and~1607555, the Sloan Foundation, 
and the Leverhulme Trust. 
Computing resources were provided by the University of Chicago Research Computing Center.

\bibliographystyle{apsrev4-1}
\bibliography{paper}

\end{document}